\documentclass[onecolumn,12pt,journal,a4paper,draftclsnofoot]{IEEEtran}

\usepackage{lineno}
\usepackage{setspace}
\usepackage{xcolor}

\usepackage[fleqn]{amsmath}
\usepackage{wasysym,calc,capt-of,ifthen}
\usepackage{amsfonts,txfonts,amssymb,verbatim,relsize}
\usepackage{subfigure}
\usepackage{graphicx}
\usepackage{psfrag,color}
\usepackage{fullpage}
\usepackage{epsfig}
\usepackage{amssymb}
\usepackage{amsmath}
\usepackage{pst-all,color,pseudocode}

\newtheorem{definition}{Definition}
\newtheorem{lemma}{Lemma}
\newtheorem{proposition}{Proposition}
\newtheorem{theorem}{Theorem}
\newtheorem{example}{Example}

\newcommand{\nin}{\not\in}

\newcommand{\tmtextit}[1]{{\itshape{#1}}}
\newcommand{\tmfloatcontents}{}
\newlength{\tmfloatwidth}
\newcommand{\tmfloat}[5]{
  \renewcommand{\tmfloatcontents}{#4}
  \setlength{\tmfloatwidth}{\widthof{\tmfloatcontents}+1in}
  \ifthenelse{\equal{#2}{small}}
    {\ifthenelse{\lengthtest{\tmfloatwidth > \linewidth}}
      {\setlength{\tmfloatwidth}{\linewidth}}{}}
    {\setlength{\tmfloatwidth}{\linewidth}}
  \begin{minipage}[#1]{\tmfloatwidth}
    \begin{center}
      \tmfloatcontents
      \captionof{#3}{#5}
    \end{center}
  \end{minipage}}

\begin{document}

%\begin{frontmatter}
  \title{\Large Combinatiorial Algorithms for Wireless Information Flow}
  
   \author{\large Javad Ebrahimi and Christina Fragouli
   \\
  EPFL\\
   Lausanne, Switzerland\\
   }
  % }
  
%  \address{}

\date{}
\maketitle

\begin{abstract}

A long-standing open question in information theory is to characterize
the unicast capacity of a wireless relay network.  The difficulty
arises due to the complex signal interactions induced in the network,
since the wireless channel inherently broadcasts the signals and 
there is interference among transmissions.  
Recently, Avestimehr, Diggavi and Tse proposed a linear  deterministic model that takes into account the shared nature of wireless channels, focusing on the signal interactions rather than the background noise. They
generalized the min-cut max-flow theorem for graphs to networks of
deterministic channels and proved that the capacity can be achieved
using information theoretical tools. They showed that the value of the
minimum cut is in this case the minimum rank of all the 
adjacency matrices describing source-destination cuts. 
%However, since
%there exists an exponential number of cuts, identifying the capacity
%through exhaustive search becomes infeasible.

In this paper,%\footnote{Parts of this work appeared in \cite{soda}.}, 
we develop a polynomial time algorithm that discovers
the relay encoding strategy to achieve the min-cut value in 
linear deterministic (wireless) networks, for the case of a unicast
connection. Our algorithm crucially uses a notion of linear
independence between channels to calculate the capacity in polynomial time. 
Moreover, we can achieve the capacity by using very simple
one-symbol processing at the intermediate nodes, thereby constructively
yielding finite length strategies that achieve the unicast capacity of
the linear deterministic (wireless) relay network. 

\end{abstract}

%\end{frontmatter}

\section{Introduction}

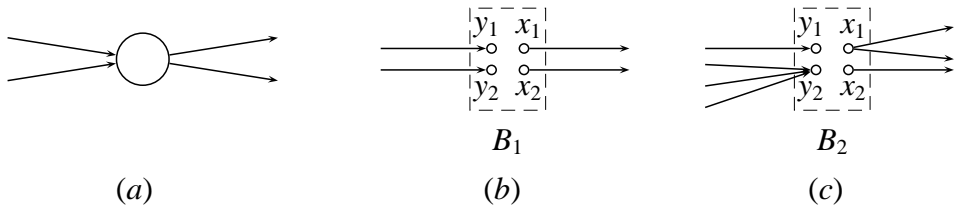
\begin{figure}[h]
\begin{center}
%\begin{scriptsize}
\psset{unit=0.028in}

\begin{pspicture}(80,-15)(130,25)

\psset{linewidth=0.2mm}

\cnode(40,11){5}{x}%{$B_1$}

\pnode(15,15){x1}
\pnode(15,7){x2}
\ncline{->}{x1}{x}
\ncline{->}{x2}{x}
\pnode(65,15){y1}
\pnode(65,7){y2}
\ncline{->}{x}{y1}
\ncline{->}{x}{y2}

\put(35,-15){$(a)$}

% B_1
\cnode(104.5,13){1}{y_1}
\nput[labelsep=1]{90}{y_1}{\textit{$y_1\;$}}
\cnode(104.5,9){1}{y_2}
\nput[labelsep=1]{270}{y_2}{\textit{$y_2\;$}}

\cnode(110.5,13){1}{x_1}
\nput[labelsep=1]{90}{x_1}{\textit{$\;x_1$}}
\cnode(110.5,9){1}{x_2}
\nput[labelsep=1]{270}{x_2}{\textit{$\;x_2$}}

\pnode(84,13){B1}
\ncline{->}{B1}{y_1}
\pnode(84,9){B2}
\ncline{->}{B2}{y_2}

\pnode(130,13){A1}
\ncline{->}{x_1}{A1}
\pnode(130,9){A2}
\ncline{->}{x_2}{A2}

% 2 invisible nodes to make the box B_2
\pnode(107.5,17){B_2a}
\pnode(107.5,5){B_2b}
\ncbox[nodesep=.25cm,boxsize=7,linewidth=0.1mm,
linestyle=dashed]{B_2a}{B_2b}

% Label ``B_2''
\nput[labelsep=7]{270}{B_2b}{$B_1$}
\put(103,-15){$(b)$}

%===============================================
\cnode(164.5,13){1}{2y_1}
\nput[labelsep=1]{90}{2y_1}{\textit{$y_1\;$}}
\cnode(164.5,9){1}{2y_2}
\nput[labelsep=1]{270}{2y_2}{\textit{$y_2\;$}}

\cnode(170.5,13){1}{2x_1}
\nput[labelsep=1]{90}{2x_1}{\textit{$\;x_1$}}
\cnode(170.5,9){1}{2x_2}
\nput[labelsep=1]{270}{2x_2}{\textit{$\;x_2$}}

\pnode(144,13){2B1}
\ncline{->}{2B1}{2y_1}
\pnode(144,10){2B2}
\ncline{->}{2B2}{2y_2}
\pnode(144,6){2B22}
\ncline{->}{2B22}{2y_2}
\pnode(144,2){2B23}
\ncline{->}{2B23}{2y_2}

\pnode(190,11){2A1}
\ncline{->}{2x_1}{2A1}
\pnode(190,17){2A11}
\ncline{->}{2x_1}{2A11}

\pnode(190,9){2A2}
\ncline{->}{2x_2}{2A2}

% 2 invisible nodes to make the box B_2
\pnode(167.5,17){2B_2a}
\pnode(167.5,5){2B_2b}
\ncbox[nodesep=.25cm,boxsize=7,linewidth=0.1mm,
linestyle=dashed]{2B_2a}{2B_2b}

\nput[labelsep=7]{270}{2B_2b}{$B_2$}
\put(163,-15){$(c)$}

\end{pspicture}
\caption{\small (a) A node in a directed graph, (b) equivalent representation through orthogonal channels, and (c) a node in a network of deterministic channels.}
\label{fig_nodes}
%\end{scriptsize}
\end{center}
\end{figure}

Let $G=(V,E)$ denote a directed graph with unit capacity edges. We can think of  each edge of this graph as a channel {\em orthogonal} to 
all other channels, where each channel (edge) has a single input and a single output,
and can be used to send a single symbol from the input to the output (unit capacity).
We can then depict  a node with multiple incoming and outgoing edges as having 
multiple inputs and multiple outputs, as determined by its adjacent edges,
where inputs and outputs can be arbitrarily connected to each other within the node.
For example, Fig.~\ref{fig_nodes}(a) depicts a node in a directed graph, and Fig.~\ref{fig_nodes}(b)
the equivalent representation of this node.

%In particular, 
%We can think of each node as having multiple inputs and multiple outputs,
%one corresponding to each adjacent incoming and outgoing edge.

Wireless relay networks cannot be represented as graphs, due to the inherently shared nature of the wireless medium that causes
complex signal interactions.
In the wireless medium, transmissions are  broadcasted, and may be received by multiple receivers at different signal strengths   depending on path loss parameters. Moreover, there is interference between transmissions, and the signal from different nodes in the network can be received at very different power at a given receiver (high dynamic range of received signals).
The characterization of the unicast capacity of a wireless relay network has been an open problem for decades, 
mainly due to these complex signal interactions.

Recently, Avestimehr, Diggavi and Tse \cite{suhas1,suhas2} proposed a linear  deterministic network model  (we will call this ADT model) that takes into account the interactions between the signals in a wireless network, i.e., broadcasting and interference, and represents the noise by a deterministic threshold rather than a random variable. 
The symbols received below the noise threshold are discarded.  The argument is that for high Signal-to-Noise-Ratio (SNR), it is the signal interactions that will dominate the performance, and thus the capacity of the deterministic 
could be very close to that of  the noisy network.
Thus networks of deterministic channels could be used  as approximate models for  wireless networks. 

The ADT model is based on the intuition of dividing the transmitted and received signals into symbols, where each symbol is transmitted at a different power level,
and assuming that only symbols above a deterministic noise  threshold will be successfully received.
 Deterministic networks can be over over an arbitrary field $\mathbf{F}_q$.  In the following, when 
we do not explicitly specify the field, we will imply that the network operates over the binary field.
%In the following we will for simplicity describe these ideas over the binary field.

As an example, consider a point-to-point AWGN channel: $y = 2^{\alpha/2}x + z$, and assume
that input  bits  $x_1,x_2,..,x_n$ are transmitted from a node $A$, while a node $B$ observes the signal $y$.
The capacity is $\log(1 + 2^\alpha) \approx \alpha \log(2)$, assuming $z$ is unit variance noise ($\alpha$ represents the channel gain in dB scale $\alpha \leftrightarrow \lceil \log(SNR)\rceil$).
The ADT model over $\mathbf{F}_2$ in this case is obtained  by truncating the received signal and assuming that the $\alpha$ most significant bits (MSB) of $x$ are {\em always} above the deterministic noise threshold and received successfully at node $B$. 
The parameter $\alpha$ captures the path loss and determines how many of the MSB bits of $x$ are received at $y$.

\begin{figure}[!t]
\begin{center}
%\begin{scriptsize}
\psset{unit=0.028in}
\begin{pspicture}(30,-10)(175,80)
\psset{linewidth=0.2mm}

% Source
\pnode(50,45){i_1}
\cnode(50,41){1}{x_1}
\nput[labelsep=1]{90}{x_1}{\textit{$x_1$}}
\pnode(50,39){i_2}
\cnode(50,37){1}{x_2}
\nput[labelsep=1]{270}{x_2}{\textit{$x_2$}}
\pnode(50,33){i_3}

% Box at the Source
\ncbox[nodesep=.25cm,boxsize=6,linewidth=0.1mm,
linestyle=dashed]{i_1}{i_3}

% Label ``Source''
\nput[labelsep=11]{180}{i_2}{\textit{S}}

% A_1

\cnode(74.5,66){1}{y_1}
\nput[labelsep=1]{90}{y_1}{\textit{$y_1\;$}}
\cnode(74.5,62){1}{y_2}
\nput[labelsep=1]{270}{y_2}{\textit{$y_2\;$}}

\cnode(80.5,66){1}{x_3}
\nput[labelsep=1]{90}{x_3}{\textit{$\;x_3$}}

% 2 invisible nodes to make the box A_1
\pnode(77.5,70){A_1a}
\pnode(77.5,58){A_1b}
\ncbox[nodesep=.25cm,boxsize=7,linewidth=0.1mm,
linestyle=dashed]{A_1a}{A_1b}

% Label ``A_1''
\nput[labelsep=7]{90}{A_1a}{$A_1$}

% A_2
\cnode(74.5,13){1}{y_3}
\nput[labelsep=1]{270}{y_3}{\textit{$y_3\;$}}

\cnode(80.5,13){1}{x_4}
\nput[labelsep=1]{270}{x_4}{\textit{$\;x_4$}}

% 2 invisible nodes to make the box A_2
\pnode(77.5,17){A_2a}
\pnode(77.5,5){A_2b}
\ncbox[nodesep=.25cm,boxsize=7,linewidth=0.1mm,
linestyle=dashed]{A_2a}{A_2b}

% Label ``A_2''
\nput[labelsep=7]{270}{A_2b}{$A_2$}

% B_1
\cnode(114.5,66){1}{y_6}
\nput[labelsep=1]{90}{y_6}{\textit{$y_6\;$}}

\cnode(120.5,66){1}{x_5}
\nput[labelsep=1]{90}{x_5}{\textit{$\;x_5$}}

% 2 invisible nodes to make the box B_1
\pnode(117.5,70){B_1a}
\pnode(117.5,58){B_1b}
\ncbox[nodesep=.25cm,boxsize=7,linewidth=0.1mm,
linestyle=dashed]{B_1a}{B_1b}

% Label ``B_1''
\nput[labelsep=7]{90}{B_1a}{$B_1$}

% B_2
\cnode(114.5,13){1}{y_7}
\nput[labelsep=1]{270}{y_7}{\textit{$y_7$}}

\cnode(120.5,13){1}{x_6}
\nput[labelsep=1]{90}{x_6}{\textit{$x_6$}}
\cnode(120.5,9){1}{x_7}
\nput[labelsep=1]{270}{x_7}{\textit{$x_7$}}

% 2 invisible nodes to make the box B_2
\pnode(117.5,17){B_2a}
\pnode(117.5,5){B_2b}
\ncbox[nodesep=.25cm,boxsize=7,linewidth=0.1mm,
linestyle=dashed]{B_2a}{B_2b}

% Label ``B_2''
\nput[labelsep=7]{270}{B_2b}{$B_2$}

% Destination
\pnode(150,45){i_4}
\cnode(150,41){1}{y_8}
\nput[labelsep=1]{90}{y_8}{\textit{$y_8$}}
\pnode(150,39){i_5}
\cnode(150,37){1}{y_9}
\nput[labelsep=1]{270}{y_9}{\textit{$y_9$}}
\pnode(150,33){i_6}

% Box at the Destination
\ncbox[nodesep=.25cm,boxsize=6,linewidth=0.1mm,
linestyle=dashed]{i_4}{i_6}

% Label ``Destination''
\nput[labelsep=11]{0}{i_5}{\textit{D}}

% Connection between the Source and A_1
\ncline{->}{x_1}{y_1}
\ncline[linewidth=0.7mm]{->}{x_2}{y_2}

% Connection between the Source and A_2
\ncline{->}{x_1}{y_3}

% Connection from A_1
\ncline[linewidth=0.7mm]{->}{x_3}{y_6}
\ncline{->}{x_3}{y_7}

% Connection from A_2
\ncline{->}{x_4}{y_6}
\ncline{->}{x_4}{y_7}

% Connection between B_1 and the Destination
\ncline[linewidth=0.7mm]{->}{x_5}{y_9}

% Connection between B_2 and the Destination
\ncline{->}{x_6}{y_8}
\ncline{->}{x_7}{y_9}

\end{pspicture}
\caption{\small An example of a  linear binary deterministic network.}
\label{fig_net1}
%\end{scriptsize}
\end{center}
\end{figure}
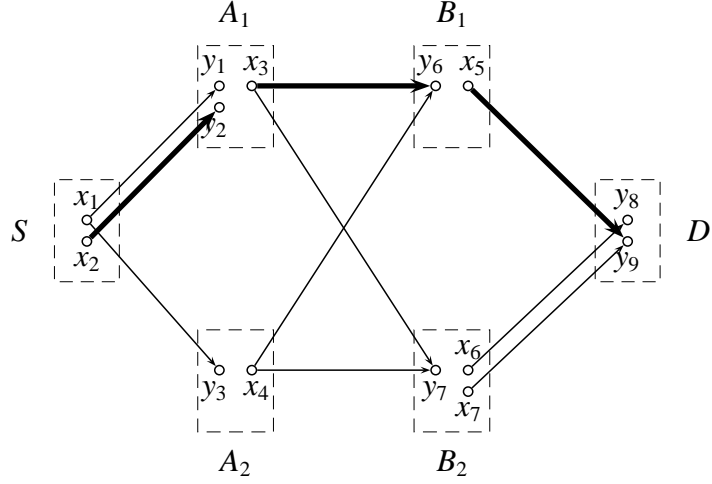

When broadcasting, each receiver node $B_i$ will receive the $m_i$ MSB  from the transmitted  bits  $x_1,x_2,..,x_n$, with $0\leq m_i \leq n$.
For example, when in Fig.~\ref{fig_net1} node $S$ transmits, node $A_1$ receives both the transmitted bits, while node $A_2$ receives only the MSB that was transmitted with the higher power. The difference between the bit index at the transmitter and the bit index at
the receiver represents path loss.

Interference in the ADT model is modeled through bit-wise binary addition,
unlike Gaussian networks, where interfering signals are added through regular addition. 
%\footnote{All  operations in our employed models are performed over the binary field $\mathbb{F}_2$.}
In Fig.~\ref{fig_net1} the  bit $y_6$ equals the binary addition (xor) of bits $x_3$ and $x_4$.
Again, the signal from different nodes in the network can be received at  different power at a given receiver. 
For example, node $D$ observes at $y_9$ the xor of $x_5$ and $x_7$, i.e., the MSB from node $B_1$ and the $2^{nd}$ MSB from node $B_2$. The generalization over an arbitrary field $\mathbf{F}_q$ is straightforward, by substituting binary addition with addition over  
 $\mathbf{F}_q$.

In the ADT model, unlike graphs, channels are no longer orthogonal. Each  input might be connected to multiple outputs
belonging in different nodes, and the relationship between these inputs and outputs is determined by a  set of linear equations. In Fig.~\ref{fig_net1},
the channel between the nodes $A_1$, $A_2$ and $B_1$, $B_2$ can be described through the equations
 $y_6=y_7=x_3+x_4$.
A generic node of deterministic channel networks is depicted in  Fig.~\ref{fig_nodes}(c).
Loosely speaking, in deterministic networks, we
can have Linear Dependence (LD) relationships between edges (we will make this precise in the following section), 
even though these edges might not be adjacent. For example, in Fig.~\ref{fig_net1}, 
the edges $(x_3,y_6)$ and $(x_4,y_7)$
are linearly dependent.  This makes challenging the task of calculating the min-cut value between a source-destination (S-D) pair
and of identifying the node operations.

 Avestimehr, Diggavi and Tse
generalized the min-cut max-flow theorem for graphs to networks of
deterministic channels and proved that the capacity can be achieved
using information theoretical tools. They showed that the value of the
minimum cut is in this case the minimum rank of all the
adjacency matrices describing source-destination cuts.
For example, in Fig.~\ref{fig_net1} the minimum cut value equals

\begin{equation}\nonumber
\text{rank}
\begin{array}{cc}
& \begin{array}{cc}  y_6 & y_7  \\
  \end{array}\\
\begin{array}{c}  x_3   \\ x_4\\
  \end{array}
&  \left( \begin{array}{cc}
 1 & 1  \\
 1 & 1  
\end{array} \right)\end{array}=1.
\end{equation}

%========================================================================
Note that there exists an exponential number of cuts, and thus identifying the capacity
through exhaustive search becomes infeasible.
In this paper, we develop a constructive polynomial-time algorithm which allows to efficiently calculate the min-cut value between a $S-D$ pair, and to achieve this value using simple operations at relay nodes.  

To construct our algorithm, it is easy to see that, attempting to directly extend the Ford-Fulkerson (FF) algorithm \cite{Ford}, or other path-augmenting algorithms developed for graphs, is not straightforward. Indeed, 
assume that  in Fig.~\ref{fig_net1} we have at a first iteration identified the path highlighted in bold.
The FF algorithm may attempt to employ the path consisting of the edges $(x_1,y_3)$, $(x_4,y_7)$, $(x_7,y_9)$, which in fact is vertex disjoint
(excluding the S, D nodes) from the already identified path. However, because edges $(x_3,y_6)$ and $(x_4,y_7)$ are LD, this path cannot bring innovative information to the destination; in fact, the min-cut value in this network equals one. Given that channels can interact in multiple ways, it is not clear that  a polynomial algorithm does exist.

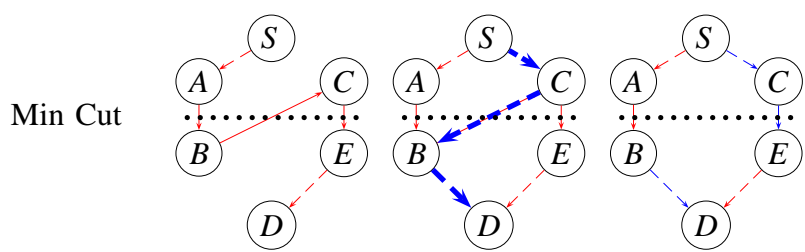
\begin{figure}[!hc]
\begin{center}
\psset{unit=0.015in}
\begin{pspicture}(0,-10)(190,90)
\psset{linewidth=0.1mm}
 
%Fig 1
%Vertices
 
\cnodeput(30,70){S1}{$S$}
\cnodeput(5,55){A1}{$A$}
\cnodeput(5,30){B1}{$B$}
\cnodeput(55,55){C1}{$C$}
\cnodeput(55,30){E1}{$E$}
\cnodeput(30,5){D1}{$D$}
 
% Edges
\ncline[linestyle=dashed,linecolor=red]{->}{S1}{A1}
\ncline[linecolor=red]{->}{A1}{B1}
\ncline[linecolor=red]{->}{B1}{C1}
\ncline[linecolor=red]{->}{C1}{E1}
\ncline[linestyle=dashed,linecolor=red]{->}{E1}{D1}
 
%Fig 2
%Vertices
 
\cnodeput(105,70){S2}{$S$}
\cnodeput(80,55){A2}{$A$}
\cnodeput(80,30){B2}{$B$}
\cnodeput(130,55){C2}{$C$}
\cnodeput(130,30){E2}{$E$}
\cnodeput(105,5){D2}{$D$}
 
% Edges
\ncline[linestyle=dashed,linecolor=red]{->}{S2}{A2}
\ncline[linecolor=red]{->}{A2}{B2}
\ncline[linecolor=red]{->}{B2}{C2}
\ncline[linecolor=red]{->}{C2}{E2}
\ncline[linestyle=dashed,linecolor=red]{->}{E2}{D2}
\ncline[linestyle=dashed, linewidth=0.7mm,linecolor=blue]{->}{S2}{C2}
\ncline[linestyle=dashed, linewidth=0.7mm,linecolor=blue]{->}{C2}{B2}
\ncline[linestyle=dashed, linewidth=0.7mm,linecolor=blue]{->}{B2}{D2}

%Fig 3
%Vertices
 
\cnodeput(180,70){S3}{$S$}
\cnodeput(155,55){A3}{$A$}
\cnodeput(155,30){B3}{$B$}
\cnodeput(205,55){C3}{$C$}
\cnodeput(205,30){E3}{$E$}
\cnodeput(180,5){D3}{$D$}
 
% Edges
\ncline[linestyle=dashed,linecolor=red]{->}{S3}{A3}
\ncline[linecolor=red]{->}{A3}{B3}
\ncline[linecolor=blue]{->}{C3}{E3}
\ncline[linestyle=dashed,linecolor=red]{->}{E3}{D3}
\ncline[linestyle=dashed,linecolor=blue]{->}{S3}{C3}
\ncline[linestyle=dashed,linecolor=blue]{->}{B3}{D3}
 
%MINCUT

\pnode(0,42.5){i_1}
\pnode(60,42.5){i_2}
 
\pnode(75,42.5){i_3}
\pnode(135,42.5){i_4}
 
\pnode(150,42.5){i_5}
\pnode(210,42.5){i_6}
 
\ncline[linestyle=dotted, linewidth=0.7mm]{-}{i_1}{i_2}
\ncline[linestyle=dotted, linewidth=0.7mm]{-}{i_3}{i_4}
\ncline[linestyle=dotted, linewidth=0.7mm]{-}{i_5}{i_6}
\put(-60,40){Min Cut}

\end{pspicture}
\caption{\small Correcting a ``mistake'' in the Ford-Fulkerson algorithm can be thought of as ``rewiring paths''.}
\label{mincut}
\end{center}
\end{figure}

Even in regular graphs, the number of cuts between an S-D pair is exponentially large.
However, polynomial time algorithms do exist in that case. One way to understand this is by observing that,
in the FF algorithm for example,  we are allowed to make ``mistakes'' when selecting a path, where a mistake in this case
is when a path crosses a minimum cut more than once. The strength of the algorithm comes from the fact that such mistakes
can be ``corrected'', by allowing to use employed edges in the opposite direction. 
What these corrections do is effectively ``rewiring'' already identified partial-paths. For example in Fig.~\ref{mincut},
a first iteration identifies the path that uses the edges $AB$, $BC$ and $DE.$ This path crosses a min-cut twice.
A subsequent iteration can use edge $CA$ in the opposite direction to find a new S-D path. This amounts to, no longer using edge $BC$
and having two rewired paths: The first part of the first path  arrives 
at node B, and is then complemented by the second path from B to D.
The second path arrives from S to E, and from E to D is complemented by the first path. 
%assume that at a first iteration we have identified path A.

In deterministic networks, we cannot avoid making ``mistakes'' when selecting which paths to use, where now a mistake amounts to using the wrong edges between a set of linearly dependent edges; thus, to find a polynomial time algorithm, we need to put in place some simple mechanisms to ``correct'' such mistakes. As we will see in following sections, now using edges in opposite directions is no longer sufficient or helpful; we may in fact have to ``jump'' across nodes, and change the inputs or outputs employed by already identified paths. The interesting and surprising point is that, there exists a method to perform such corrections in polynomial time, and thus, no ``mistake'' is catastrophic.

We close this section by noting that in \cite{suhas1}, it was  observed that to study coding strategies and achievable rates, 
we can reduce an arbitrary network into a layered network, through a  time-expansion technique,   
with asymptotically no rate-loss. Thus in this paper we will also focus our attention in layered networks,
which will be defined formally in the next section. %However, we underline that the result applies to arbitrary networks,
%with cycles, etc.

This paper is based on the work in \cite{soda}. The algorithm in
\cite{soda} was presented over binary fields. Moreover, the proof of
the algorithm presented in \cite{soda} applies under some assumptions
on the structure of the linear dependency between inputs and
outputs. In this paper, we provide a simple modification of 
 \cite{soda} that holds with no assumptions on the linear dependency
of the channels. Moreover, we  present the algorithm  
over an arbitrary finite field $\mathbf{F}_q$. The paper in \cite{soda}
was followed up by a very nice connection with matroids 
and the development of alternative algorithms for this problem \cite{matroid_allerton}.

%\begin{Example}

This paper is organized as follows. Section~\ref{sec1} introduces our notation and basic definitions.
%and gives an example where constructive use of interference gives significant
%benefits. 
Section~\ref{sec2} describes our algorithm, provides a number of examples, and proves that it
identifies a minimal value cut. Section \ref{sec3} concludes the paper.

\section{Model and Definitions}\label{sec1}

In this section, we start by defining the layered 
deterministic network model for a unicast connection over a network. 
%As we
%already mentioned, this model is a generalization of the previous model. We refer
%to it as the deterministic model but the reader should be aware of the
%difference.
%x
\begin{definition}
  {\em (Layered Deterministic Network).} A layered deterministic network model over a finite field $\mathbf{F}_q$, consists of a set of nodes and a set of channels (or edges) with the following properties:
 \begin{enumerate} 
  \item Each node consist of two  sets, the set of inputs and the set of outputs of the node. We will generally denote inputs using the variable $x$ and outputs using the variable $y$. We will denote by $A(x)$ and $A(y)$ respectively, the node where input $x$
 (output $y$) belongs to.  Let $I_{total}$ be the  total number of inputs  in the network and
$O_{total}$ the total number of outputs in the network.

  \item The nodes of the deterministic model are partitioned in  parts. Each part
  is called a layer of the network.  We assume that each layer has at most $M$ nodes, and denote by $V_i$ the set of nodes in layer $i$. 
The layers are labeled by $i=1, 2, \ldots,
  \text{$\Lambda$},$ where $\Lambda$ is the number of layers.
  
  \item Layer 1 and layer $\Lambda$ each has only one node in it. The node of the first
  layer is called ``source node'' and is denoted by ${S}$ and the node of the last
  layer is called ``receiver node'' and is denoted by ${D}$. The source node has only outputs and the receiver node has only inputs.
  
  \item Each channel is a link between an input of a node in layer $i$ to an
  output of a node in layer $i + 1$ where $1 \leq i \leq \Lambda - 1$. A fixed nonzero value over a finite field $\mathbf{F}_q$ is associated with each link.
  
  \item Let  $\mathbf{x}$ denote a vector that collects all inputs in layer $i$, and  $\mathbf{y}$ a vector that collects all outputs in the next layer $i+1$. Then these vectors are connected through a given linear transformation over  $\mathbf{F}_q$, i.e.,
%$\begin{equation} \nonumber
$\mathbf{y}=\mathbf{T}\mathbf{x}$,
%$\end{equation}
where each nonzero value in the transformation matrix $\mathbf{T}$ corresponds to a channel and its associated  value.\hfill{$\square$}
\end{enumerate}
\end{definition}

We can define a transformation matrix between an arbitrary subset of inputs and outputs in adjacent layers.
Let  $V$ be a subset of all inputs in layer $i$ and $W$ be a subset of all
outputs in layer $i + 1$ (for simplicity we do not include the $i$ indices).

\begin{definition}
  {\em (Transfer Matrix).} We define $\mathbf{T} (V, W)$ to be the matrix whose rows
  are labeled with the elements of $V$, the columns with the elements of $W$
  and the entry $(v, w)$ is nonzero 
  if and only if there is a channel between input
  $v$ and output $w$. $\mathbf{T}(V, W)$ is called the transfer matrix between $V$ and
  $W$.\hfill{$\square$}
\end{definition}

%Note that the transfer matrix is not necessarily square, because for
%example multiple edges in $U$ might share the same $x_i$ or $y_j$. Sometimes
%it will be convenient to associate the transfer matrix not with a set of
%links but rather with a set of inputs and outputs. We will then use the
%notation $\mathbf{T} (U_x, U_y)$ to specify the row and column set of
%$\mathbf{T}$. 
We will describe the extension of a given
transformation matrix $\mathbf{T} (V, W)$ by adding a row corresponding to
an input $x \nin V$ as $\mathbf{T} (\{V, x \}, W)$ and the extension
by adding both a row corresponding to an input $x$ and a column corresponding to an output $y$ (not already contained in
$V$ and $W$) as $\mathbf{T} (\{V, x \}, \{W, y \})$.

The maximum information $S$ can send to $D$ depends on the minimum cut value in the network, defined as follows.
\begin{definition} \label{def_cut}
  {\em (Cut and Cut-Value)} By an $S - D$ cut $\mathcal{V}_C$ we mean a partition of
  the nodes into two parts $\mathcal{V}_1$ and $\mathcal{V}_2$ in such a way that $S \in \mathcal{V}_1$ and $T
  \in \mathcal{V}_2$. We define the value of $\mathcal{V}_C$ to be 
$\text{rank}\mathbf{T} (A_1,
  A_2) \log_2 q$, where $\text{rank}$ refers to matrix rank,  $A_1$ is the set of all inputs in the nodes in $\mathcal{V}_1$, $A_2$
  is the set of all outputs of the nodes is $\mathcal{V}_2$, and $q$ is the size of the employed finite field. The minimum cut value equals $min_{\mathcal{V}_C}\text{rank}\mathbf{T} (A_1,
  A_2) \log_2 q$, where the minimization is over all $S-D$ cuts. \hfill{$\square$}
\end{definition}

%In {\cite{suhas1,suhas2}} it is proved that information rate equal the
%min-cut value can be sent from $S$ to $D$. The provided proof information theoretical tools.
%\iffalse
We will sometimes distinguish between a {\em layer-cut} and a {\em cross-cut}.
There exist exactly $\Lambda - 1$ layer cuts, one between every two
consecutive layers. For example, the $j$-layer cut is $\mathcal{V}_1 = V_1 \cup
\ldots \cup V_j$ and $\mathcal{V}_2 = V_{j + 1} \cup \ldots \cup
V_{\Lambda}$ for $1 \leq j \leq \Lambda - 1$. A cross-cut involves several
layers. The transfer matrix for a cross-cut is block diagonal, with the nodes
in each layer belonging in a different block.
%\fi

Next, we will define the notion of linear dependency between channels. 
\begin{definition}
  {\em (LI and LD Channels).} Suppose that $H$ is a subset of channels
  between layers $i$ and $i + 1$. Let $V$ be the set of all inputs that are
  the head of a channel in $H$ and $W$ be the set of tails of these channels.
  We say $H$ is a set of Linearly Independent (LI) channels  if $\text{rank} \mathbf{T} (V,
  W) = |H|$. Otherwise we say $H$ is a set of Linearly Dependent (LD) channels.
  \label{def11}\hfill{$\square$}
\end{definition}

%Each output of a node (apart from the source node) can generally be a function  of inputs of that node.
%  In this paper we select these functions to be linear functions. The coefficients of the linear equations are not given beforehand and they
%  should be determined. Our algorithm  selects these coefficients. Essentially, our algorithm will simply  map a subset of the inputs of each node to a subset of that nodes outputs. Equivalently, it selects at each node a (potentially empty) set of used inputs and outputs. 

Our algorithm will send information from $S$ to $D$ using $S-D$ paths, defined in the following. Through every path $S$ sends one symbol over $F_q$ to $D$.
\begin{definition}
  {\em ($S-D$ Path).} 
An $S-D$ path  is a disjoint set of edges $(e_1, e_2,
\ldots, e_{\Lambda - 1})$ where $e_1$ starts from $S$, $e_{\Lambda - 1}$
finishes at $D$, and $e_i$ finishes at the same node where edge $e_{i + 1}$
starts. All $S-D$ paths have the same length $\Lambda - 1$, because of the
structure of the layered network. \hfill{$\square$}
\end{definition}

Essentially, selecting paths amounts to appropriately selecting sets of inputs $V$ and outputs $W$  to be used in each layer.
To ensure that the information send through different paths can be decoded at the destination we need to use linearly independent (LI) paths, defined as follows.
\begin{definition}\label{def_LI_path}
  {\em (LI-Paths).} Suppose that $\mathcal{P}$ is a set of $S-D$ paths. We say these paths
  are linearly independent if and only if the set of edges of these paths in
  every layer form a set of linearly independent edges. \hfill{$\square$}
\end{definition}

Note that each $x$ and $y$ can take part in at most one of the LI
paths; in this case we will say that it is  {\em used} by that path. That is, we will  say
that a channel input $x$ is used, if there exists a path that
uses  a channel $(x, y')$ for some $y'$. Similarly, we will say that a channel
output $y$ is used, if there exists a path that uses a channel
$(x', y)$ for some $x'$.

\section{The Unicast Algorithm}\label{sec2}
%==================================================

\subsection{Main idea}
%==================================

In our algorithm, we
will find linearly independent paths one after another, in iterations.
 The
first iteration identifies a path $\mathcal{P}_1$. This is always possible if
the source is connected to the destination, otherwise the capacity is zero.
Each subsequent iteration identifies an additional path such that all selected
paths are LI (as by definition \ref{def_LI_path}).
For example at iteration $K + 1$, the algorithm takes as input the LI paths
$\mathcal{P} =\{ \mathcal{P}_1, \ldots, \mathcal{P}_K \}$ and attempts to find
path $\mathcal{P}_{K + 1}$ such that the paths $\{ \mathcal{P}_1, \ldots,
\mathcal{P}_{K + 1} \}$ are also LI (as by definition \ref{def_LI_path}).
 Each iteration finishes once we reach the destination.
The algorithm stops when an iteration cannot complete, at which point the algorithm outputs the set of identified LI paths $\mathcal{P}$.

To find a new path, we start from the source and select one channel at each layer
until we reach the destination if possible. At each layer we need to select a valid channel, in the sense that it is linearly independent from the set of the channels of the  identified paths in that layer in previous iterations.  
A main tool that we use to achieve this is that we allow the algorithm to perform 
some type of ``rewiring'' inside one layer at a time. Assume for example that we have
$K + 1$ ``partial'' paths from the source to nodes in layer $i$, and $K$
``partial'' paths from nodes from layer $i + 1$ to the destination. Rewiring
refers to that we change the mapping between the starting and finishing paths by changing
the channels we employ,
while still preserving LI across the $i$-layer cut. 
These rewiring are achieved through two functions, the
 $L$-function  and the $\phi$-function,  which
we describe in detail later.
  
Note that, instead of selecting channels (or paths), we can equivalently think of our algorithm as appropriately selecting a subset of inputs and outputs to be used in each layer. Each node internally simply maps each of its used inputs to a used output (the specific mapping is not important).  That is, selecting $K$  paths amounts to selecting  $K$ inputs $U_x$ at each layer $i$ and $K$ outputs $U_y$ at the corresponding layer $i+1$ such that the transfer matrix $\mathbf{T}_i=\mathbf{T}(U_x,U_y)$ is full rank for each $i$.

Each of the LI paths that the algorithm outputs will be used to convey
an independent symbol over the field $\mathbf{F}_q$ from the source to the destination. 
Let $\mathbf{x}$ collect the $K$ used outputs of the source and 
$\mathbf{y}$ collect the $K$ used inputs of the receiver.
The overall transfer matrix $\mathbf{T} = \mathbf{T}_1 \cdot \mathbf{T}_2  \ldots  \mathbf{T}_{\Lambda -1} $ is full rank and therefore $\mathbf{x}$ can be recovered at the receiver by solving the system of linear equations \[ \mathbf{y}=\mathbf{T}  \mathbf{x} =\mathbf{T}_1 \cdot \mathbf{T}_2  \ldots  \mathbf{T}_{\Lambda -1}\mathbf{x}.\]
That is, although 
we send one symbol through each path,  due to the linear combining the deterministic model imposes, the receiver will still need to solve equations to retrieve the
data.   By the choice of paths, that is, by selecting at each
node the edges we use to collect and transmit information, we preserve the
``degrees of freedom'',  the number of independent linear equations the
receiver decodes.  
%\\
%\textcolor{blue}{
%\begin{example}
%An example where we show some LI paths?
%\end{example}
%}

% As we already
%mentioned the algorithm is path augmenting and operates in iterations. 

\subsection{Algorithm Description}
%==================================

Assume we are at iteration $K + 1$, that takes as input the LI paths
$\mathcal{P} =\{ \mathcal{P}_1, \ldots, \mathcal{P}_K \}$ and attempts to find
path $\mathcal{P}_{K + 1}$ such that the paths $\{ \mathcal{P}_1, \ldots,
\mathcal{P}_{K + 1} \}$ are also LI (as by definition \ref{def_LI_path}).

During this iteration, we {\em explore} nodes, starting from the source node $S$.
We will use the terminology of exploring a node $A$ to indicate that we
have found a path from $S$ to  $A$ (LI from the paths in $\mathcal{P}$)
and attempt to continue this path from node $A$ to $D$ in order to
complete $\mathcal{P}_{K + 1}$. Note that which input $y_i \in A$ we use to reach the
node $A$ does not play a role; to explore a node it is sufficient that we arrive
at it using any of its inputs. Once we reach a node, we {\em mark} the node as
visited, and attempt to explore all edges emanating from it, as potential
candidates for the path $\mathcal{P}_{K + 1}$. We use an indicator variable 
$\mathcal{M}$
with values $\{ \texttt{T}, \texttt{F} \}$, to mark whether a node or edge has
been explored $( \texttt{T})$ or not $( \texttt{F})$. We need explore (according
to operations to be defined) each node during each iteration at most once, and
we will do that calling a function $E_A$. 
Exploring a node reduces to
exploring all unused inputs that it contains; exploring an input is achieved
by calling a function $E_x$. Each edge may be explored during each iteration
multiple times, for reasons we will explain in the following, but no more than
a finite number of times. This ensures that each iteration terminates after a
finite number of steps.

Assume that we have found a partial path $\mathcal{P}_{K+1}$ from $S$ to a node $A$ in the $i$-layer
and 
we explore input $x_i\in A$, with the goal of extending the path $\mathcal{P}_{K+1}$ to the $i+1$-layer.
Let $U =\{(x_i, y_j)\}$
denote the set of $K$ used edges in the $i$-layer cut, $U_x$ and $U_y$ denote
the set of used inputs and outputs respectively, and $\mathbf{T} (U_x, U_y)$
be the $K \times K$ full rank transformation matrix associated with $U$.
We describe the steps we take to explore a specific input
in the following. We illustrate these steps through a number of examples in Section~\ref{sec_examples}.

\subsection*{Steps in exploring input $x_i$ at node $A$}

\begin{enumerate}
%\tmtextbf{1.} 
\item \label{C1} If $x_i \in U_x$, i.e., $x_i$ is already used by a path,  do
nothing. Note that although node $A$ will be marked as explored ($\mathcal{M}
(A) = \texttt{T}$), this particular $x_i \in A$ will not be marked ($\mathcal{M}
(x_i) = \texttt{F}$ will remain).
%\tmtextbf{2.} 
\item \label{C2} If $x_i \nin U_x$,  i.e., $x_i$ is not used,  then for each $y_j$,
such that the channel $(x_i, y_j)$ exists, we distinguish two cases.
\begin{enumerate}
\item  \label{C2b} $y_j \notin U_y$, i.e., $y_j$ is not used. Consider the $(K + 1) \times (K + 1)$ 
matrix $\mathbf{T} (\{U_x, x_i \}, \{U_y, y_j \})$ associated with the used
edges and the new edge $(x_i, y_j)$. We again consider two cases.
\begin{enumerate}
\item \label{C2bi} If the matrix $\mathbf{T} (\{U_x, x_i \}, \{U_y, y_j \})$ is not full rank, 
do nothing. 

%In this case there
%exists at least one edge $(x_i, y_k)$ with $y_k \in U_y$. Indeed, otherwise if
%the last row of $\mathbf{T}$ had zeroes everywhere apart the last element, we
%would have that \[\det ( \mathbf{T} (\{U_x, x_i \}, \{U_y, y_j \})) = \det (
%\mathbf{T} (U_x, U_y)) \neq 0.\] Since such an edge exists, case~(\ref{C2a}) will at some point be executed, and as we will see there are
%no additional actions needed.

\item \label{C2bii} If the matrix $\mathbf{T} (\{U_x, x_i \}, \{U_y, y_j \})$ is
full rank, use edge $(x_i, y_j)$ to go to node $A (y_j)$. If this node has not
been visited before, we attempt to continue from node $A (y_j)$ by calling the
function $E_A (G, \mathcal{P}, \mathcal{M}, A (y_j))$.\\
Additionally, for each $y_k \in U_y$, with $A (y_k) = A (y_j)$,
perform what we call the {$\phi$-function}. The idea is that, in this
case there exists a path from the source to the destination identified during
a previous iteration that goes through node $A(y_j)$. This path uses an edge $(x_k,
y_k) \in U$ to reach node $A (y_j)$. We can then use our newly identified
partial path that uses the edge $(x_i, y_j)$ to reach from the source the node
$A (y_j)$, and ``connect'' this new partial path with the existing partial
path from $A(y_j)$ to destination. Thus, we have the opportunity to again perform
rewirings and visit new nodes.

More precisely, the $\phi$-function performs the
following.
Remove from the matrix $\mathbf{T} (\{U_x, x_i \}, \{U_y, y_j \})$ the
column corresponding to  $y_k$ with $A (y_k) = A (y_j)$. Let 
\[\mathbf{C}
\triangleq \mathbf{T} (\{U_x, x_i \}, \{U_y, y_j \}- \{y_k\})\] denote the
resulting $(K + 1) \times K$ matrix. Consider each of the $K$ square submatrices 
of $\mathbf{C}$ resulting by deleting each of the first $K$ rows. Let
$\mathbf{C}_m$ denote the submatrix resulting from deleting row $x_m$, i.e., 
\[\mathbf{C}_m
\triangleq \mathbf{T} (\{U_x, x_i \}- \{x_m\}, \{U_y, y_j \}- \{y_k\}).\]
If
$\mathbf{C}_m$ is not full rank, do nothing. If it is full rank, perform a
rewiring of the existing $K$ paths using $\mathbf{C}_m$. 
 If $A (x_m)$ is not marked as visited, explore $A (x_m)$. If $A (x_m)$ is
marked as visited, then explore input $x_m$ even if it is marked.  
Note that the $\phi$-function may be executed at most as many times
as the number of outputs in that layer, and thus when it is executed, at most $K$ already visited
inputs might be revisited. Examples \ref{example_2}-\ref{example_4} illustrate the use of the $\phi$ function.

\end{enumerate}

\item \label{C2a} $y_j \in U_y$, i.e.,  $y_j$ is used. 
We can then not immediately use the channel $(x_i, y_j)$, unless we perform some rewiring. 
This rewiring is captured by what we call the $L_x$-function. 
This function will be executed at most once for every input. To ensure that,
we keep in the algorithm for each input an indicator variable 
$\mathcal{ML}$ with values $\{ \texttt{T}, \texttt{F} \}$. 

The $L_x$-function operates as follows. Consider the extended
transformation matrix $\mathbf{T} (\{U_x, x_i \}, U_y)$. Define $L_{x_i}
\subseteq U_x$ to be the smallest subset of $U_x$, of size $|L_{x_i} | = s
\leq K$, such that the matrix $\mathbf{T} (\{L_{x_i},
x_i \}, U_y)$ has rank $s$. Using proposition~\ref{prop1} this set can be
identified in polynomial time. Proposition~\ref{prop2} proves that removing
any one of the rows of $\mathbf{T} (\{L_{x_i},
x_i \}, U_y)$  still results in a full rank matrix.
Equivalently, removing any row of $\mathbf{T} (U_x, U_y)$ corresponding to a
$x_k \in L_{x_i}$, and substituting it with the row corresponding to $x_i$,
results in a full rank matrix, that can be used to rewire the paths identified
in the previous iterations. That is, using Proposition~\ref{prop0}, we can use
the row $\mathbf{T} (x_i,U_y)$ to substitute any of the already employed 
$\mathbf{T} (x_k,U_y)$, $x_k \in L_{x_i}$ that are LD
with $x_i$ row, while still maintaining the same number of paths as identified
from the previous iterations. We will then be left with a partial path
arriving at the node $A (x_k)$, and we can attempt to use any of the available
$x$'s in this node to proceed. We now distinguish to subcases:
\begin{enumerate}
\item $A (x_k)$ is already marked as
explored. In this case we will not visit this node again. However, we will explore input $x_k$, although this input might have been explored before. Note that, at each execution of the $L_x$ function, at most $K$ inputs will be re-examined.
\item   $A(x_k)$ is not marked, i.e., during this iteration we visit
this node for the first time. Then the algorithm explores this node. Additionally, if 
 there exists a path identified during a previous
iteration that utilizes (at the previous layer) an output $y'$ at node $A(x_k)$ 
we will execute on this node the $\phi$-function that we described previously.
%Example~\ref{example_4} illustrates such a situation. 
\end{enumerate}
Examples \ref{example_1} and \ref{example_4} illustrate the use of the $L_x$-function.

%\tmfloat{h}{small}{figure}{\includegraphics{hope-last so far-2-7.eps}}{Left:
%Paths before rewiring, Middle Paths after algorithm runs Right: Identified cut
%if node B3
%\ \ \ \ \ \ \ \ \ and B4 had no outgoing edges}

\end{enumerate}
\end{enumerate}

\iffalse
\fbox{
%\subsection*{L-function}
Given an input $x_i$, the L-function performs the following:
\begin{itemize}
\item Find $L_{x_i}$. 
\item For each $x \in L_{x_i}$:
\begin{itemize}
\item  if $A(x)$ has not been visited before, 
use $U_x-\{x\}\cup\{x_i\}$ as the current set of used inputs and explore $A$. As this is the first time we explore node $A(x)$, we will also perform the $\phi$-function, that we will describe later on.
\item If $A(x)$ has been visited but $x_k$ has not been visit, explore $x_k$.
\item Otherwise do nothing.
\end{itemize}
\end{itemize}
}
\fi
%In this step we encountered some LD relationships. For each input $x \in U$ we
%will keep a set of associated inputs $R_x$, whose purpose we will discuss
%later; in this set we will add all inputs that occurred in the same set
%$L_{x_i}$ as $x$ for any $x_i$.

The previous steps are the main ingredients of two recursive functions  $E_A$ and $E_X$ 
that implement our proposed algorithm and  are
%To find the paths, our algorithm uses two recursive functions 
summarized in Table~\ref{tabl1}. 
%$E_A$ takes as input the network, the node $A$ from
%where we start (the first time we call the function $A = S$), the set of
%identified paths in the previous and current iteration (summarized in a
%structure $\mathcal{P}$), and the set of already marked (visited) nodes and
%edges (described by the function $\mathcal{M}$). We assume that $\mathcal{P}$
%contains the information for $U$, $U_x$ and $U_y$. The function $E_A$ returns
%true if it reaches the destination and false if it fails.
 The  first function, $E_x$,  checks if we can
continue from a current input $x$  to reach the destination by a sequence of
channels which are linearly independent to the previous identified paths. 
The
input of this function is the network, a family of identified paths and
already visited nodes and current input. It returns true if there is a
sequence of channels with the described properties and returns false,
otherwise.  
The second function, $E_A$, does a similar job as the first function except that it
works for the current node instead of the current channel. So, as one might
guess, this function, essentially, calls the first function for all of its
inputs and if none of them returns true, it also returns false.
We  illustrate the algorithm steps through a number of 
examples in Section~\ref{sec_examples}.% \textcolor{blue}{To add more examples.}

%-If the input $x_m$ is marked, this implies that, there exists an input
%$x_{\ell} \nin U_x$ and an associated edge that could not be used due to LD,
%and $L_{x_{\ell}}$ contained $x_m$. Since $x_m$ will now be removed from the
%set of used edges, it may be possible to use now edges that we could not
%before. To achieve that, we need to be able to examine again all inputs that
%belong in $\{L_{x_{\ell}} \}$, although they have been examined before. This
%is why, for every input $x_{\ell}$ we have kept track of the set of related
%inputs $R_{x_{\ell}}$. When $x_m$ is removed from the set of used inputs, the
%set $R_{x_m}$ will contain all $x$ that potentially need to be examined again.
%We can thus unmark all these inputs, and attempt to find a path starting from
%$x_m$. Note that the $\phi$-function will be executed at most $M$ times, where
%$M$ is the maximum number of vertices in a layer.

\begin{small}
\begin{table*}[!t]
\begin{center}
\begin{pseudocode}[doublebox]
%{\{(T,F)\}=E_A}{  G,\mathcal{P},\mathcal{M}, A}
{set of functions $E_A$ and $E_x$}{.}%G,,\mathcal{P},\mathcal{M},A or x}
\{(\text{T},\text{F})\}=E_A(G,\mathcal{P},\mathcal{M}, A)\\
\IF  \mathcal{M}(A)==\text{T} \quad \RETURN{\text{F}}
\ELSE \\
\BEGIN
\mathcal{M}(A)=\text{T}\\
U\GETS \{\text{used edges in $L(A)$-layer cut}\}, \; U_x\GETS \{x_i \in U\}, \;  U_y\GETS \{y_j \in U\},\;
\mathcal{X}\GETS  \{x_i \in A\}\\
\forall x_i \in \mathcal{X},\quad 
\IF x_i \notin U_x
\AND M(x_i)==F 
\AND E_x(G,P,M,x_i)==\text{T}\quad \RETURN{\text{T}}\\
\END\\
\RETURN {\text{F}}\\
\\
\\
%The second function%%%%%%%%%%%%%%%%%%%%%%
\{(T,F)\}=E_x(  G,\mathcal{P},\mathcal{M}, x_i)\\
%\BEGIN %at the end
\IF  \mathcal{M}(x_i)==\text{T} \quad \RETURN{\text{F}}
\ELSE \\
\BEGIN
\mathcal{M}(x_i)=\text{T}\\ %% mark it
\forall y_j: (x_i,y_j)\in E\\
\IF  y_j \in U_y
\BEGIN
\IF ML(x_i)==F \text{\small $\%$ (we perform this function only once per input)}\\
\BEGIN
  ML(x_i)=T \quad \\
L_x=\text{FindL($\mathbf{T}(\{U_x,x_i\},U_y)$)}\\
%\forall  x_k \in \{L_x\}, R_{x_k} \gets \{ R_{x_k} \cup L_x\}\\
\forall \quad  x_k \in L_x \\
%\mbox{ with }  (x_k,y_k) \in U, \quad  y_k \in L_y  \\
\BEGIN
\text{Match($\mathbf{T}(\{L_x-x_k,x_i\},{L}_y)$)}\\
\text{Update($\mathcal{P}$)}\\
\IF \mathcal{M}(A(x_k))==\text{F}\\
\BEGIN
\forall \quad y_k\; \in \; A(x_k)\;\;
\text{ \small perform $\phi$-function($y_k$) {\tiny (see description following)}}\\
\IF  E_A( G,\mathcal{P},\mathcal{M}, A(x_k))==\text{T} \quad \RETURN{\text{T}}
\END
\ELSE %\mathcal{M}(A(x_k))=\text{T}, \;\;  % visit node A(x_k)
\BEGIN
\IF \mathcal{M}(x_k)==\text{T} 
\BEGIN
 \text{Set} \;\; \mathcal{M}(x_k)=\text{F}\\
\IF  E_x( G,\mathcal{P},\mathcal{M}, x_k)==\text{T} \quad \RETURN{\text{T}}
\END 
\END\\
%\ELSE
%\BEGIN % visit node A(x_k)
%\IF  E_A( G,\mathcal{P},\mathcal{M}, A(x_k))=\text{T} \quad \RETURN{\text{T}}\\
%\END\\
\text{Restore($\mathcal{P}$)}\\
\END
\END
\END\\
\\ %***************** Second case ******************************************
\IF  y_j \notin U_y
\BEGIN
\IF \text{rank}(\mathbf{T}(\{U_x,x_i\},\{U_y,y_j\}))=1+\text{rank}(\mathbf{T}(U_x,U_y))\\
\BEGIN
\IF A(y_j)==\text{Destination} \quad
\RETURN{\text{T}}
\\
\IF \mathcal{M}(A(y_j))==F\\
\BEGIN
%%%%%\IF  E_A(G,\mathcal{P},\mathcal{M}, A(y_j))=\text{T} \quad \RETURN{\text{T}}\\ 
\forall y_k\in U_y \text{ with } A(y_k)==A(y_j) \AND (x_k,y_k) \in U \\
\textcolor{black}{\phi-\text{function}(y_k):}
\BEGIN 
\forall x_k  \in U_x\\
\IF \text{$\mathbf{T}(\{U_x-x_k,x_i\},\{U_y-y_k,y_j\})$) is full rank}\\
 U_y=\{U_y-y_k,y_j\},U_x=\{U_x-x_k,x_i\}\\
\text{Update($\mathcal{P}$)}\\
%\IF \mathcal{M}(A(x_i))=\text{F}
\IF \mathcal{M}(A(x_k))== \text{F}\\
\BEGIN 
\forall \quad y_\ell\; \in \; A(x_k)\;\;
\text{ \small perform $\phi$-function($y_\ell$)}\\
\IF  E_A(G,\mathcal{P},\mathcal{M}, A(x_k))==\text{T} \quad \RETURN{\text{T}}
\END
\ELSE\\
\IF \mathcal{M}(x_k)==\text{T} \\
\BEGIN
 \text{Set} \;\; \mathcal{M}(x_k)=\text{F}\\
\IF  E_x( G,\mathcal{P},\mathcal{M}, x_k)==\text{T} \quad \RETURN{\text{T}}
\END \\
%E_x( G,\mathcal{P},\mathcal{M}, x_i)=\text{T} \quad \RETURN{\text{T}}\\
\text{Restore($\mathcal{P}$)}\\
\END\\
\IF  E_A(G,\mathcal{P},\mathcal{M}, A(y_j))=\text{T} \quad \RETURN{\text{T}}\\ 
\END
\END
\END
\END\\
\RETURN{\text{F}}
\label{alg_LIF}
\end{pseudocode}
\end{center}
\caption{\label{tabl1}\small  The  functions $E_A(\cdot)$ and $E_x(\cdot)$ are executed by the algorithm at each explored node or edge. The function ``Match($\mathbf{T}$)'' finds a perfect matching in the bipartite graph defined by matrix $\mathbf{T}$ as
described in proposition~\ref{prop0}. 
The function ``FindL($\mathbf{T}$)'' finds the smallest set of rows that are LD with the last row of $\mathbf{T}$ as described in proposition~\ref{prop1}. The function ``Update($\mathcal{P}$)'' keeps track of the current wiring of identified paths (which may change either by the execution of the match function, or by the $\phi$-fucntion),
while ``Restore($\mathcal{P}$)'' restores $\mathcal{P}$ to the value before the last update. 
%The set $R_{x_k}$, initialized to the empty set at the beginning of each iteration, keeps for every input $x_k$ the current set of identified inputs  that are in a LD relationship with $x_k$. 
The labeling function $M(\cdot)$ equals T if a node (or edge) is already explored, in which case it is not explored again, and F otherwise. The labeling function $ML(\dot)$ similarly keeps rack if the $FindL$ function has been executed for an input; this function is executed at most once per input. The $\phi$ function is essentially executed every time we visit for the first time a node where a semi-path no longer used arrives.}
\end{table*}\end{small}
\thispagestyle{empty}

\subsection*{Three Propositions Used in the Algorithm}
We here provide some useful propositions that were used in our algorithm.
The first is a
known property {\cite{Harvey}}, that allows to ``match'' inputs and outputs through LI channels, and that
 that we repeat for completeness.
\begin{proposition}
  \label{prop0}If the $K \times K$ binary matrix $\mathbf{T} (U_x, U_y)$ is
  full rank, then there exist $K$ LI edges with $x \in U_x$ and $y \in U_y$.
\end{proposition}
\begin{proof}
%\tmtextit{Proof:} 
Since $\mathbf{T} (U_x, U_y)$ is full rank matrix, it has
nonzero determinant. Now if we expand the determinant using the sum of product-
expansion, we should have at least one non-zero product and this product
corresponds to a perfect matching in the bipartite graph with adjacency matrix
$\mathbf{T} (U_x, U_y)$.%{\hspace{fill}}$\square$
\end{proof}

\begin{proposition}
  \label{prop1}Let $\mathbf{T} (U_x, U_y)$ be a full rank $K \times K$ matrix
  and $\mathbf{x}_i \triangleq \mathbf{T} (x_i, U_y)$ a vector in its span. Then, we can
  find the smallest $L_{x_i} \subseteq U_x$ of size $s = |L_{x_i} | \leq K$
  such that $\text{rank} ( \mathbf{T} (\{L_{x_i}, x_i \}, U_y)) = \text{rank}
  ( \mathbf{T} (L_{x_i}, U_y)) = s$ using $O (K^3)$ operations.
\end{proposition}
\begin{proof}
%\tmtextit{Proof:} 
Since the matrix $\mathbf{A}\triangleq \mathbf{T} (U_x, U_y)$ is full rank,
there exists a unique vector  $\mathbf{c}$ such that 
$\mathbf{x}_i=\mathbf{c}\mathbf{A}$. Solve these equations to find $\mathbf{c}$.
$L_{x_i}$ are the indices corresponding to nonzero values in $\mathbf{c}$.
% For each $x_j \in U_x$, if $\text{rank} ( \mathbf{T} (\{U_x
%- x_j, x_i \}, U_y)) = K$ then $x_j \in L_{x_i}$. The idea is the vectors in
%$\{U_x - x_j, x_i \}$ will have rank less than $K$ only if they include all
%the vectors in $L_{x_i}$ and $x_i$. This algorithm takes $O (K^4)$ operations.
%{\hspace{fill}}$\square$
\end{proof}

\begin{proposition}
  \label{prop2}Let $L_{x_i}$ be the smallest subset of $U_x$, $|L_{x_i} | =
  s$, such that $\text{rank} ( \mathbf{T} (\{L_{x_i}, x_i \}, U_y \}) = \text{rank} (
  \mathbf{T} (L_{x_i}, U_y)) = s$. Then for each $x_j \in L_{x_i}$, $\text{rank} (
  \mathbf{T} (\{L_{x_i} - x_j, x_i \}, U_y \}) = s$.
\end{proposition}
\begin{proof}
%\tmtextit{Proof:} 
Consider the vectors $\mathbf{x}_j \triangleq \mathbf{T}
(x_j, U_y)$.  From minimality of $L_{x_i}$, $\mathbf{x}_i = \sum_{j \in
L_{x_i}}\alpha_j \mathbf{x}_j$ with $\alpha_j \neq 0$, otherwise, we could have found a smaller set to
replace $L_{x_i}$. Thus, for any $x_k \in L_{x_i}$,
\begin{eqnarray*}
  \mathbf{x}_i = \beta_k \mathbf{x}_k + \sum_{x_j \in L_{x_i}, \hspace{0.75em} j \neq
  k} \beta_j \mathbf{x}_j 
\end{eqnarray*}
for some nonzero coefficients $\beta$'s over the finite field.
Since the vectors $\{ \mathbf{x}_j \}$ with $x_j \in L_{x_i}$ are LI, and
since given $\mathbf{x}_i$ and all other $\mathbf{x}_j$ apart $\mathbf{x}_k$
we can still retrieve $\mathbf{x}_k$, the matrix $\mathbf{T} (\{L_{x_i} - x_j,
x_i \}, U_y \})$ has full rank.  
%\hfill{$\square$}
\end{proof}

%Sometimes we may not be given a set  of LI edges, but instead, a set of
%$U_x$ and $U_y$ which we are asked to ``match'' if possible to create LI edges
%$(x, y)$ with $x \in U_x$ and $y \in U_y$. 

%%%%%%%%%%%%%%%%%%%%%%%%%%%%%%%%%%%%%%%%%%%%%%%%%%%%%%%%%%%%%%%%%%%%%%%%%%%%%%%%%%%%%

\begin{figure*}[!t]
\begin{center}
\input{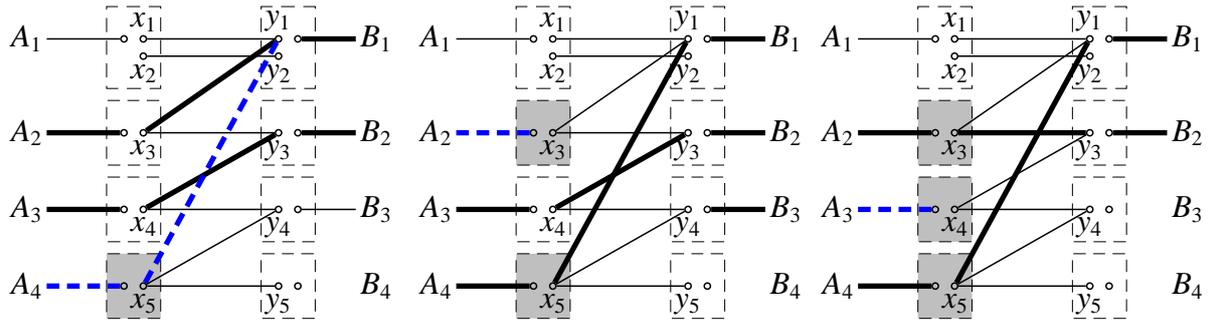}
\caption{\small Assume that  bold  depict edges in paths $\mathcal{P}_1$, $\mathcal{P}_2$ identified through previous iterations.
At iteration $3$ a partial path $\mathcal{P}_3$ arrives at node $A_4$, and we explore  the edge $(x_5,y_1)$. We perform rewiring using the $L_{x}$ function. Left: marked nodes and paths before the $L_{x}$ function. Middle: marked nodes and paths when substituting $x_4$ with $x_5$.  Right: marked nodes and paths when substituting $x_3$ with $x_5$.}
\label{ex1}
\end{center}
\end{figure*}

\subsection{Examples}\label{sec_examples}
%======================================

%\subsection*{Example 1}
\begin{example}{\em Exploring an input and the $L_x$-function.}\label{example_1}
Consider the layer cut in the left Fig.~\ref{ex1} and assume that, during iterations 1 and 2, we have identified the two LI paths $\mathcal{P}_1$ and $\mathcal{P}_2$ that use the bold edges in the figure. Thus, 
\[
%\begin{aligned}
 U=\{(x_3,y_1), (x_4,y_3)\}, 
\quad U_x=\{x_3,x_4\},  \quad
 U_y=\{y_1,y_3\},
 \quad \mathbf{T}(U_x,U_y)=
\begin{array}{cc}
& \begin{array}{cc}  y_1 & y_3  \\
  \end{array}\\
\begin{array}{c}  x_3   \\ x_4\\
  \end{array}
&  \left( \begin{array}{cc}
 1 & 1  \\
 0 & 1  
\end{array} \right)\end{array}.
%\begin{array}{ccc}
%     & y_1 & y_3 \\
%x_3& 1 & 1\\
%x_4& 0 & 1\\\end{array} 
%\end{aligned}
\]

%\begin{figure}[!hc]
%\begin{center}
%\input{Figures/fig1}
%\caption{Left: paths before rewiring, Right: paths after the algorithm runs, new wiring of the edges used.}
%\label{fig1}
%\end{center}
%\end{figure}

In iteration 3, assume  that we  reach node $A_4$. We mark this node as visited,
and  examine the channel input $x_5$. 
There are three possible edges we need to explore: $\{(x_5,y_1),\; (x_5,y_4),\; (x_5,y_5)\}$.

\begin{itemize}
\item We first  examine the edge $(x_5,y_1)$.  This is depicted in the left Fig.~\ref{ex1}.
%\begin{itemize}
% 	\item 
Since $y_1\in U_y$, we are at step $(2-b)$ of the algorithm.  We thus consider the matrix
\[
\mathbf{T}(\{U_x,x_5\},U_y)=
\begin{array}{cc}
& \begin{array}{cc}  y_1 & y_3  \\
  \end{array}\\
\begin{array}{c}  x_3   \\ x_4\\ x_5\\
  \end{array}
&  \left( \begin{array}{cc}
 1 & 1  \\
 0 & 1  \\
 1 & 0 \\
\end{array} \right)\end{array},
 \]

and find the set $L_{x_5}=\{x_3,x_4\}$.

We can attempt to substitute each of the $x\in L_x$ with $x_5$.
	\begin{itemize}
 		\item  If we substitute $x_3$, we mark $A_2$ and find another matching: $\{(x_5,y_1),(x_4,y_3)\}$. This is depicted in the middle Fig.~\ref{ex1}. Since it is the first time we visit node $A_2$, and since there is path arriving at it, we will perform the $\phi$-function at this node. We will not describe these steps here, see for such a case example \ref{ex4}. We then call $E(G,\mathcal{P},\mathcal{M},A_2)$. Assume this function returns \texttt{F}, i.e., fails to find a path to the destination. We restore the original path matching and continue.

	 	\item If we substitute $x_4$: We mark $A_3$ and find another matching: $\{(x_5,y_1),(x_3,y_3)\}$.  This is depicted in the right Fig.~\ref{ex1}.  We again perform the $\phi$-function at node $A_3$, bit described in this example.
	 We then call $E(G,\mathcal{P},\mathcal{M},A_3)$. Again assume it fails to find a path to the destination. We restore the original path matching and continue.

	\end{itemize}

\begin{figure*}[!t]
\begin{center}
\input{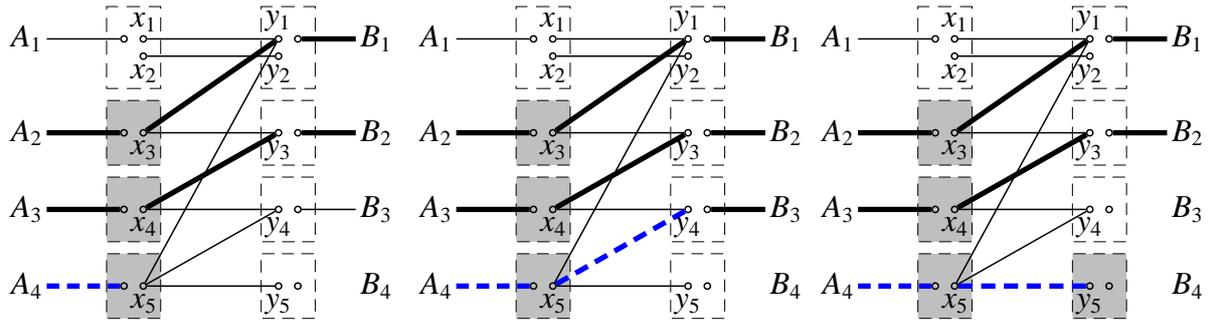}
\caption{\small Continuing from the example in Fig. \ref{ex1}. Failing to use the edge $(x_5,y_1)$, we will next explore the edges $(x_5,y_4)$, and $(x_5,y_5)$.   Left: marked nodes and paths. Middle: marked nodes and paths  when exploring  the edge $(x_5,y_4)$. Right: marked nodes and paths when exploring  the edge $(x_5,y_5)$.}
\label{ex1-1}
\end{center}
\end{figure*}

\item We proceed  with $(x_5,y_4)$, as depicted in the middle Fig.~\ref{ex1-1}. Since $y_4 \notin U_Y$, we examine the rank of the matrix
\[
\mathbf{T}(\{U_x,x_5\},\{U_y,y_4\})=
\begin{array}{cc}
& \begin{array}{ccc}  y_1 & y_3  & y_4 \\
  \end{array}\\
\begin{array}{c}  x_3   \\ x_4\\ x_5\\
  \end{array}
&  \left( \begin{array}{ccc}
 1 & 1 & 0 \\
 0 & 1 & 1\\
 1 & 0 & 1\\
\end{array} \right)\end{array}.
 \]

Because $\text{rank}(\mathbf{T}(\{U_x,x_5\},\{U_y,y_4\}))=\text{rank}(\mathbf{T}(U_x,U_y))=2$ we are at step $(2-a-i)$ of the algorithm, and  	we do not need to take any actions.%,  the algorithm does not do anything.

	\item Finally, for the edge  $(x_5,y_5)$, with
$y_5 \notin U_y$,  we examine the rank of the matrix 
\[ \mathbf{T}(\{U_x,x_5\},\{U_y,y_5\})=
\begin{array}{cc}
& \begin{array}{ccc}  y_1 & y_3  & y_5 \\
  \end{array}\\
\begin{array}{c}  x_3   \\ x_4\\ x_5\\
  \end{array}
&  \left( \begin{array}{ccc}
 1 & 1 & 0 \\
 0 & 1 & 1\\
 0 & 0 & 1\\
\end{array} \right)\end{array}
 \]
	%\item 
Since $\text{rank}(\mathbf{T}(\{U_x,x_5\},\{U_y,y_5\}))=\text{rank}(\mathbf{T}(U_x,U_y))+1=3$, we are at step $(2-a-ii)$ of the algorithm, and  
we can use the edge   $(x_5,y_5)$ in the path $\mathcal{P}_3$. We thus mark node $B_4$ as visited and continue from there.

That is, we update $\mathcal{P}$, and we call the function
 $E(G,\mathcal{P},\mathcal{M},B_4)$. 
Note that since $\mathcal{P}_1$ and $\mathcal{P}_2$ do not use node $B_4$, we will not perform the $\phi$-function at this node.
This is depicted in right Fig. \ref{ex1-1}.
%Assume it fails to find a path to the destination.
%\end{itemize}

%\begin{itemize}

%\end{itemize}
\hfill{$\square$}
\end{itemize}

\end{example}

We next provide two examples for the $\phi$-function.
\begin{example} {\em First example for $\phi$-function.}
\label{example_2}

\begin{figure*}[!t]
\begin{center}
\input{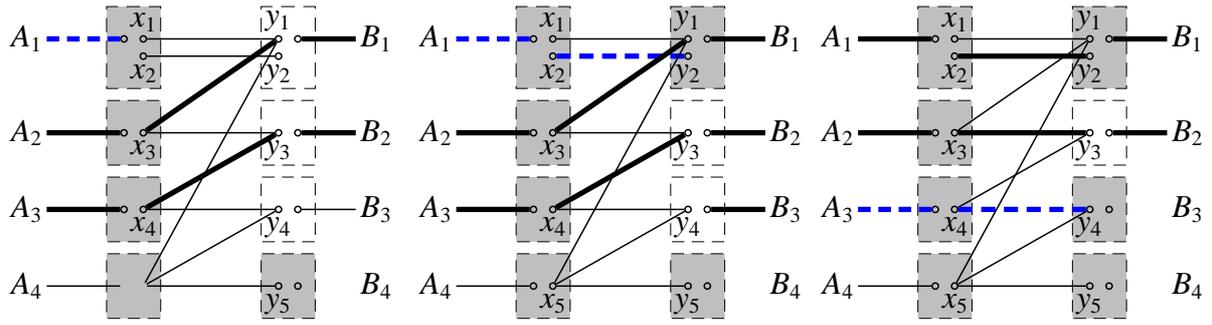}
\caption{\small Continuing from the example in Figs. \ref{ex1} and \ref{ex1-1}.}
\label{ex2}
\end{center}
\end{figure*}

Continuing the previous example, assume that we have failed to find a path when exploring $A_4$. Suppose that the algorithm continues
and suppose that, through some different path, we  reach and mark node $A_1$, as depicted in the left Fig.~\ref{ex2} (we maintain the marked nodes from the previous algorithm steps during this iteration).
We will now explore inputs $x_1$ and $x_2$.

Assume we start by edge  $(x_2,y_2)$. We can use this edge to reach and mark $B_1$, as depicted in the middle  Fig.~\ref{ex2}. Since this is the first time we visit node $B_1$,  we will perform the $\phi$-function.  

\[
\mathbf{T}(U_x\cup\{x_2\},U_y\cup \{y_2\} -\{y_1\})=
\mathbf{T}(\{x_2,x_3,x_4\},\{y_2,y_3\})=
\begin{array}{cc}
& \begin{array}{cc}  y_1 & y_3  \\
  \end{array}\\
\begin{array}{c}  x_3   \\ x_4\\ x_5\\
  \end{array}
&  \left( \begin{array}{cc}
 1 & 1  \\
 0 & 1  \\
 0 & 1 \\
\end{array} \right)\end{array}.
\]

 Consider the transfer matrix where we remove the output $y_1$, and use the inputs $\{x_2,x_3,x_4\}$ and the outputs $\{y_2,y_3\}$.
Both submatrices $\mathbf{T}(\{x_2,x_3\},\{y_2,y_3,\})$  and $\mathbf{T}(\{x_2,x_4\},\{y_2,y_3,\})$ are full rank, and thus we can explore inputs $x_4$ and $x_3$ respectively. We will here describe the steps when selecting the submatrix  $\mathbf{T}(\{x_2,x_3\},\{y_2,y_3,\}$.
%, the steps in the other case are very similar.

We find a matching for $\mathbf{T}(\{x_2,x_3\},\{y_2,y_3,\}$, as depicted in the right  Fig.~\ref{ex2}, and proceed to examine input $x_4$. Note that since node $A_3=A(x_4)$ is already marked, we do not need to explore it again.
We observe that we can use the edge $(x_4,y_4)$, and thus we mark node $B_3$ and we can further proceed from there.
\hfill{$\square$}
\end{example}

%\subsection*{Example 2}
\begin{example} {\em Second example for $\phi$-function.}\label{example_3}
%=======================================================
\begin{figure*}[!t]
\begin{center}
\psset{unit=0.02in}
\begin{pspicture}(0,10)(300,74)
\psset{linewidth=0.2mm}

%%%%%%%%%%%%%%%%%%%%%%%%%%%%%%%%%%
%%%%%%%%%%%%%%%%%%%%%%%%%%%%%%%%%%
% A_1
\pnode(27.5,73.5){A_1a}
\pnode(27.5,60){A_1b}
\ncbox[fillcolor=lightgray, fillstyle=solid,nodesep=.2cm,boxsize=7,linewidth=0.1mm,linestyle=dashed]{A_1a}{A_1b}

\pnode(5,69){i_1}
\cnode(25,69){1}{a_1}

\pnode(5,64.5){i_11}
\cnode(25,64.5){1}{a_11}

\cnode(30,69){1}{b_1}
\nput[labelsep=1]{90}{b_1}{\textit{$x_1$}}
\cnode(30,64.5){1}{b_2}
\nput[labelsep=1]{270}{b_2}{\textit{$x_2$}}

% 2 invisible nodes to make the box A_1

\nput[labelsep=1]{180}{i_1}{$A_1$}

% A_2
% 2 invisible nodes to make the box A_2
\pnode(27.5,49){A_2a}
\pnode(27.5,40){A_2b}
\ncbox[nodesep=.2cm,boxsize=7,linewidth=0.1mm,
linestyle=dashed]{A_2a}{A_2b}
\pnode(5,44.5){i_2}

\cnode(25,44.5){1}{a_3}

\cnode(30,44.5){1}{b_3}
\nput[labelsep=1]{270}{b_3}{\textit{$x_3$}}

\nput[labelsep=1]{180}{i_2}{\textit{$A_2$}}

% A_3
% 2 invisible nodes to make the box A_3
\pnode(27.5,29){A_3a}
\pnode(27.5,20){A_3b}
\ncbox[nodesep=.2cm,boxsize=7,linewidth=0.1mm,
linestyle=dashed]{A_3a}{A_3b}
\pnode(5,24.5){i_3}

\cnode(25,24.5){1}{a_5}

\cnode(30,24.5){1}{b_5}
\nput[labelsep=1]{270}{b_5}{\textit{$x_4$}}

\nput[labelsep=1]{180}{i_3}{\textit{$A_3$}}

% B_1

% 2 invisible nodes to make the box B_1

\pnode(67.5,73.5){B_1a}
\pnode(67.5,60){B_1b}
\ncbox[fillcolor=lightgray, fillstyle=solid,nodesep=.2cm,boxsize=7,linewidth=0.1mm, linestyle=dashed]{B_1a}{B_1b}

\pnode(85,69){i_5}

\cnode(65,69){1}{c_1}
\nput[labelsep=1]{90}{c_1}{\textit{$y_1$}}
\cnode(65,64.5){1}{c_2}
\nput[labelsep=1]{270}{c_2}{\textit{$y_2$}}

\cnode(70,69){1}{d_1}

\nput[labelsep=1]{0}{i_5}{\textit{$B_1$}}

% B_2
\pnode(85,44.5){i_6}

\cnode(65,44.5){1}{c_3}
\nput[labelsep=1]{270}{c_3}{\textit{$y_3$}}

\cnode(70,44.5){1}{d_3}

% 2 invisible nodes to make the box B_2
\pnode(67.5,49){B_2a}
\pnode(67.5,40){B_2b}
\ncbox[nodesep=.2cm,boxsize=7,linewidth=0.1mm,
linestyle=dashed]{B_2a}{B_2b}

\nput[labelsep=1]{0}{i_6}{\textit{$B_2$}}

% B_3
\pnode(85,24.5){i_7}

\cnode(65,24.5){1}{c_5}
\nput[labelsep=1]{270}{c_5}{\textit{$y_4$}}

\cnode(70,24.5){1}{d_5}

% 2 invisible nodes to make the box B_3
\pnode(67.5,29){B_3a}
\pnode(67.5,20){B_3b}
\ncbox[nodesep=.2cm,boxsize=7,linewidth=0.1mm,
linestyle=dashed]{B_3a}{B_3b}

\nput[labelsep=1]{0}{i_7}{\textit{$B_3$}}

%invisible nodes connections
\ncline[linewidth=0.7mm]{-}{i_1}{a_1}
\ncline[linewidth=0.7mm,linestyle=dashed,linecolor=blue]{-}{i_11}{a_11}
\ncline[linewidth=0.7mm]{-}{i_2}{a_3}
\ncline[linewidth=0.7mm]{-}{i_3}{a_5}
\ncline[linewidth=0.7mm]{-}{d_1}{i_5}
\ncline[linewidth=0.7mm]{-}{d_3}{i_6}
\ncline[linewidth=0.7mm]{-}{d_5}{i_7}
%\ncline{-}{d_7}{i_8}

% Connection between A_1 and B_1

\ncline[linewidth=0.7mm]{-}{b_1}{c_1}
\ncline{-}{b_1}{c_3}
\ncline[linewidth=0.7mm,linestyle=dashed,linecolor=blue]{-}{b_2}{c_2}

% Connection from A_2
\ncline{-}{b_3}{c_1}
\ncline[linewidth=0.7mm]{-}{b_3}{c_3}
\ncline{-}{b_3}{c_5}
% Connection from A_3
\ncline{-}{b_5}{c_3}
\ncline[linewidth=0.7mm]{-}{b_5}{c_5}

% Connection from A_4
\ncline{-}{b_7}{c_1}
\ncline{-}{b_7}{c_5}
\ncline{-}{b_7}{c_7}

% RIGHT IMAGE

% A_1
% 2 invisible nodes to make the box A_1
\pnode(133.5,73.5){Ab_1a}
\pnode(133.5,60){Ab_1b}
\ncbox[fillcolor=lightgray, fillstyle=solid, nodesep=.2cm,boxsize=7,linewidth=0.1mm,
linestyle=dashed]{Ab_1a}{Ab_1b}
\pnode(111,69){ib_1}
\cnode(131,69){1}{ab_1}

\pnode(111,64.5){ib_11}
\cnode(131,64.5){1}{ab_11}

\cnode(136,69){1}{bb_1}
\nput[labelsep=1]{90}{bb_1}{\textit{$x_1$}}
\cnode(136,64.5){1}{bb_2}
\nput[labelsep=1]{270}{bb_2}{\textit{$x_2$}}

\nput[labelsep=1]{180}{ib_1}{$A_1$}

% A_2
% 2 invisible nodes to make the box A_2
\pnode(133.5,49){Ab_2a}
\pnode(133.5,40){Ab_2b}
\ncbox[nodesep=.2cm,boxsize=7,linewidth=0.1mm,
linestyle=dashed]{Ab_2a}{Ab_2b}
\pnode(111,44.5){ib_2}

\cnode(131,44.5){1}{ab_3}

\cnode(136,44.5){1}{bb_3}
\nput[labelsep=1]{270}{bb_3}{\textit{$x_3$}}

\nput[labelsep=1]{180}{ib_2}{\textit{$A_2$}}

% A_3
% 2 invisible nodes to make the box A_3
\pnode(133.5,29){Ab_3a}
\pnode(133.5,20){Ab_3b}
\ncbox[fillcolor=lightgray, fillstyle=solid, nodesep=.2cm,boxsize=7,linewidth=0.1mm,
linestyle=dashed]{Ab_3a}{Ab_3b}
\pnode(111,24.5){ib_3}

\cnode(131,24.5){1}{ab_5}

\cnode(136,24.5){1}{bb_5}
\nput[labelsep=1]{270}{bb_5}{\textit{$x_4$}}

\nput[labelsep=1]{180}{ib_3}{\textit{$A_3$}}

% B_1
% 2 invisible nodes to make the box B_1
\pnode(173.5,73.5){Bb_1a}
\pnode(173.5,60){Bb_1b}
\ncbox[fillcolor=lightgray, fillstyle=solid, nodesep=.2cm,boxsize=7,linewidth=0.1mm,
linestyle=dashed]{Bb_1a}{Bb_1b}
\pnode(191,69){ib_5}

\cnode(171,69){1}{cb_1}
\nput[labelsep=1]{90}{cb_1}{\textit{$y_1$}}
\cnode(171,64.5){1}{cb_2}
\nput[labelsep=1]{270}{cb_2}{\textit{$y_2$}}

\cnode(176,69){1}{db_1}

\nput[labelsep=1]{0}{ib_5}{\textit{$B_1$}}

% B_2
\pnode(191,44.5){ib_6}

\cnode(171,44.5){1}{cb_3}
\nput[labelsep=1]{270}{cb_3}{\textit{$y_3$}}

\cnode(176,44.5){1}{db_3}

% 2 invisible nodes to make the box B_2
\pnode(173.5,49){Bb_2a}
\pnode(173.5,40){Bb_2b}
\ncbox[nodesep=.2cm,boxsize=7,linewidth=0.1mm,
linestyle=dashed]{Bb_2a}{Bb_2b}

\nput[labelsep=1]{0}{ib_6}{\textit{$B_2$}}

% B_3
\pnode(191,24.5){ib_7}

\cnode(171,24.5){1}{cb_5}
\nput[labelsep=1]{270}{cb_5}{\textit{$y_4$}}

\cnode(176,24.5){1}{db_5}

% 2 invisible nodes to make the box B_3
\pnode(173.5,29){Bb_3a}
\pnode(173.5,20){Bb_3b}
\ncbox[nodesep=.2cm,boxsize=7,linewidth=0.1mm,
linestyle=dashed]{Bb_3a}{Bb_3b}

\nput[labelsep=1]{0}{ib_7}{\textit{$B_3$}}

%invisible nodes connections
\ncline[linewidth=0.7mm]{-}{ib_1}{ab_1}
\ncline[linewidth=0.7mm]{-}{ib_11}{ab_11}
\ncline[linewidth=0.7mm]{-}{ib_2}{ab_3}
\ncline[linewidth=0.7mm,linecolor=blue,linestyle=dashed]{-}{ib_3}{ab_5}
\ncline{-}{ib_4}{ab_7}
\ncline[linewidth=0.7mm]{-}{db_1}{ib_5}
\ncline[linewidth=0.7mm]{-}{db_3}{ib_6}
\ncline[linewidth=0.7mm]{-}{db_5}{ib_7}
%\ncline{-}{db_7}{ib_8}

% Connection between A_1 and B_1

\ncline{-}{bb_1}{cb_1}

\ncline[linewidth=0.7mm]{-}{bb_1}{cb_3}
\ncline[linewidth=0.7mm]{-}{bb_2}{cb_2}

% Connection from A_2
\ncline{-}{bb_3}{cb_1}
\ncline{-}{bb_3}{cb_3}
\ncline[linewidth=0.7mm]{-}{bb_3}{cb_5}

% Connection from A_3
\ncline{-}{bb_5}{cb_3}
\ncline{-}{bb_5}{cb_5}

% Connection from A_4
\ncline{-}{bb_7}{cb_1}
\ncline{-}{bb_7}{cb_5}
\ncline{-}{bb_7}{cb_7}

%%%%%%%%%%%%%%%%%%%%%%%%%%%%%%%
%%%%%%%%%%%%%%%%%%%%%%%%%%%%%%
%RIGHT IMAGE

% A_1

% 2 invisible nodes to make the box A_1
\pnode(237.5,73.5){Ab_1a}
\pnode(237.5,60){Ab_1b}
\ncbox[fillcolor=lightgray, fillstyle=solid, nodesep=.2cm,boxsize=7,linewidth=0.1mm,
linestyle=dashed]{Ab_1a}{Ab_1b}
\pnode(215,69){ib_1}

\cnode(235,69){1}{ab_1}

\cnode(240,69){1}{bb_1}

\pnode(215,64.5){ic_1}
\cnode(235,64.5){1}{ac_1}

\nput[labelsep=1]{90}{bb_1}{\textit{$x_1$}}
\cnode(240,64.5){1}{bb_2}
\nput[labelsep=1]{270}{bb_2}{\textit{$x_2$}}

\nput[labelsep=1]{180}{ib_1}{$A_1$}

% A_2
% 2 invisible nodes to make the box A_2
\pnode(237.5,49){Ab_2a}
\pnode(237.5,40){Ab_2b}
\ncbox[fillcolor=lightgray, fillstyle=solid, nodesep=.2cm,boxsize=7,linewidth=0.1mm,
linestyle=dashed]{Ab_2a}{Ab_2b}
\pnode(215,44.5){ib_2}

\cnode(235,44.5){1}{ab_3}

\cnode(240,44.5){1}{bb_3}
\nput[labelsep=1]{270}{bb_3}{\textit{$x_3$}}

\nput[labelsep=1]{180}{ib_2}{\textit{$A_2$}}

% A_3
% 2 invisible nodes to make the box A_3
\pnode(237.5,29){Ab_3a}
\pnode(237.5,20){Ab_3b}
\ncbox[nodesep=.2cm,boxsize=7,linewidth=0.1mm,linestyle=dashed]{Ab_3a}{Ab_3b}
\pnode(215,24.5){ib_3}

\cnode(235,24.5){1}{ab_5}

\cnode(240,24.5){1}{bb_5}
\nput[labelsep=1]{270}{bb_5}{\textit{$x_4$}}

\nput[labelsep=1]{180}{ib_3}{\textit{$A_3$}}

% B_1
% 2 invisible nodes to make the box B_1
\pnode(277.5,73.5){Bb_1a}
\pnode(277.5,60){Bb_1b}
\ncbox[fillcolor=lightgray, fillstyle=solid, nodesep=.2cm,boxsize=7,linewidth=0.1mm,
linestyle=dashed]{Bb_1a}{Bb_1b}

\pnode(295,69){ib_5}

\cnode(275,69){1}{cb_1}
\nput[labelsep=1]{90}{cb_1}{\textit{$y_1$}}
\cnode(275,64.5){1}{cb_2}
\nput[labelsep=1]{270}{cb_2}{\textit{$y_2$}}

\cnode(280,69){1}{db_1}

\nput[labelsep=1]{0}{ib_5}{\textit{$B_1$}}

% B_2
\pnode(295,44.5){ib_6}

\cnode(275,44.5){1}{cb_3}
\nput[labelsep=1]{270}{cb_3}{\textit{$y_3$}}

\cnode(280,44.5){1}{db_3}

% 2 invisible nodes to make the box B_2
\pnode(277.5,49){Bb_2a}
\pnode(277.5,40){Bb_2b}
\ncbox[nodesep=.2cm,boxsize=7,linewidth=0.1mm,
linestyle=dashed]{Bb_2a}{Bb_2b}

\nput[labelsep=1]{0}{ib_6}{\textit{$B_2$}}

% B_3
% 2 invisible nodes to make the box B_3
\pnode(277.5,29){Bb_3a}
\pnode(277.5,20){Bb_3b}
\ncbox[nodesep=.2cm,boxsize=7,linewidth=0.1mm,
linestyle=dashed]{Bb_3a}{Bb_3b}

\pnode(295,24.5){ib_7}

\cnode(275,24.5){1}{cb_5}
\nput[labelsep=1]{270}{cb_5}{\textit{$y_4$}}

\cnode(280,24.5){1}{db_5}

\nput[labelsep=1]{0}{ib_7}{\textit{$B_3$}}

%

%invisible nodes connections
\ncline[linewidth=0.7mm]{-}{ib_1}{ab_1}
\ncline[linewidth=0.7mm]{-}{ic_1}{ac_1}
\ncline[linewidth=0.7mm,linecolor=blue,linestyle=dashed]{-}{ib_2}{ab_3}
\ncline[linewidth=0.7mm]{-}{ib_3}{ab_5}
\ncline{-}{ib_4}{ab_7}
\ncline[linewidth=0.7mm]{-}{db_1}{ib_5}
\ncline[linewidth=0.7mm]{-}{db_3}{ib_6}
%\ncline[linewidth=0.7mm]{-}{db_5}{ib_7}
%\ncline{-}{db_7}{ib_8}

% Connection between A_1 and B_1

\ncline{-}{bb_1}{cb_1}
\ncline[linewidth=0.7mm]{-}{bb_1}{cb_3}
\ncline[linewidth=0.7mm]{-}{bb_2}{cb_2}

% Connection from A_2
\ncline{-}{bb_3}{cb_1}
\ncline{-}{bb_3}{cb_3}
\ncline{-}{bb_3}{cb_5}

% Connection from A_3
\ncline{-}{bb_5}{cb_3}
\ncline[linewidth=0.7mm]{-}{bb_5}{cb_5}

%\pscurve*[linestyle=dashed]{-}(200,74)(200,69)(177.5,56,75)(165,44.5)

\end{pspicture}
\caption{\small }
\label{ex3}
\end{center}
\end{figure*}
%==================================================
Consider the layer cut in Fig.~\ref{ex3}. Assume during the first three iterations we have identified the paths depicted with bold edges,
that is,
\[
U=\{(x_1,y_1), (x_3,y_3), (x_4,y_4)\}, \quad U_x=\{x_1,x_3,x_4\}, \\ 
 U_y=\{y_1,y_3,y_4\}, 
\quad \mathbf{T}(U_x,U_y)=
\begin{array}{cc}
& \begin{array}{ccc}  y_1 & y_2  & y_3 \\
  \end{array}\\
\begin{array}{c}  x_1   \\ x_2\\ x_3\\
  \end{array}
&  \left( \begin{array}{ccc}
 1 & 1 & 0 \\
 1 & 1 & 1\\
 0 & 1 & 1\\
\end{array} \right)\end{array}.
\]
During iteration $4$, we attempt to use edge $(x_2,y_2)$.
 Since node $B_1$ has not been used before, we perform the $\phi$-function. 
We thus consider the matrix
\[
\mathbf{T}( \{x_1,x_2,x_3,x_4\},\{y_2,y_3,y_4\})=
\begin{array}{cc}
& \begin{array}{ccc}  y_2 & y_3  & y_4 \\
  \end{array}\\
\begin{array}{c}  x_1   \\ x_2\\ x_3\\x_4
  \end{array}
&  \left( \begin{array}{ccc}
 1 & 1 & 0 \\
 1 & 0 & 0\\
 0 & 1 & 1\\
 0 & 1 & 1\\
\end{array} \right)\end{array}
\]
which has the full rank submatrices
$\mathbf{T}( \{x_1,x_2,x_3\},\{y_2,y_3,y_4\})$ and
$\mathbf{T}( \{x_1,x_2,x_4\},\{y_2,y_3,y_4\})$.
Using 
the $\mathbf{T}( \{x_1,x_2,x_3\},\{y_2,y_3,y_4\})$ and the matching depicted in the middle Fig.~\ref{ex3}, we  can visit node $A_3$ and explore input $x_4$. Note that since $A_3$  has not been visited before, we need perform the $\phi$-function on the node $A_3$ itself.

If instead we start by utilizing the submatrix $\mathbf{T}( \{x_1,x_2,x_4\},\{y_2,y_3,y_4\})$ and the matching depicted in the right Fig.~\ref{ex3}, we visit node $A_2$.
Again,  since $A_2$  has not been visited before, we need perform the $\phi$-function on the node $A_2$ as well.
\hfill{$\square$}
\end{example}

%=================================================================
The next example illustrates how the algorithm runs and performs rewirings across several layers.

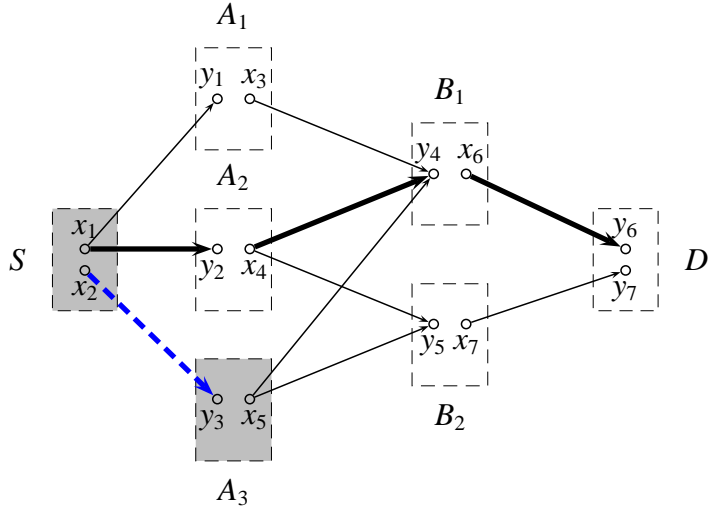
\begin{figure}[!t]
\begin{center}
%\begin{scriptsize}
\psset{unit=0.028in}
\begin{pspicture}(30,-10)(175,80)
\psset{linewidth=0.2mm}

% Source

% Box at the Source
\pnode(50,45){i_1}
\pnode(50,33){i_3}
\ncbox[fillcolor=lightgray, fillstyle=solid,nodesep=.25cm,boxsize=6,linewidth=0.1mm, linestyle=dashed]{i_1}{i_3}

\cnode(50,41){1}{x_1}
\nput[labelsep=1]{90}{x_1}{\textit{$x_1$}}

\cnode(50,37){1}{x_2}
\nput[labelsep=1]{270}{x_2}{\textit{$x_2$}}

% Label ``Source''
\pnode(50,39){i_2}
\nput[labelsep=11]{180}{i_2}{\textit{S}}
%===============================================

%  node A_1
% 2 invisible nodes to make the box A_1
\pnode(77.5,75){A_1a}
\pnode(77.5,63){A_1b}
\ncbox[nodesep=.25cm,boxsize=7,linewidth=0.1mm,
linestyle=dashed]{A_1a}{A_1b}

\cnode(74.5,69){1}{y_1}
\nput[labelsep=1]{90}{y_1}{\textit{$y_1\;$}}
%\cnode(74.5,62){1}{y_2}
%\nput[labelsep=1]{270}{y_2}{\textit{$y_2\;$}}

\cnode(80.5,69){1}{x_3}
\nput[labelsep=1]{90}{x_3}{\textit{$\;x_3$}}

% Label ``A_1''
\nput[labelsep=7]{90}{A_1a}{$A_1$}

%===========================================

% New node A_2
\cnode(74.5,41){1}{y_2}
\nput[labelsep=1]{270}{y_2}{\textit{$y_2\;$}}
%\cnode(80,42){1}{y_2}
%\nput[labelsep=1]{270}{y_2}{\textit{$y_4\;$}}

\cnode(80.5,41){1}{x_4}
\nput[labelsep=1]{270}{x_4}{\textit{$\;x_4$}}

% 2 invisible nodes to make the box A_2
\pnode(77.5,45){A_2a}
\pnode(77.5,33){A_2b}
\ncbox[nodesep=.25cm,boxsize=7,linewidth=0.1mm, linestyle=dashed]{A_2a}{A_2b}

% Label ``A_2''
\nput[labelsep=7]{90}{A_2a}{$A_2$}

%===============================
%=============================
% A_3
% 2 invisible nodes to make the box A_4
\pnode(77.5,17){A_3a}
\pnode(77.5,5){A_3b}
\ncbox[fillcolor=lightgray, fillstyle=solid,nodesep=.25cm,boxsize=7,linewidth=0.1mm,
linestyle=dashed]{A_3a}{A_3b}

\cnode(74.5,13){1}{y_3}
\nput[labelsep=1]{270}{y_3}{\textit{$y_3\;$}}

\cnode(80.5,13){1}{x_5}
\nput[labelsep=1]{270}{x_5}{\textit{$\;x_5$}}

% Label ``A_2''
\nput[labelsep=7]{270}{A_3b}{$A_3$}

%==================================================
\iffalse
% B_1
\cnode(114.5,41){1}{y_4}
\nput[labelsep=1]{90}{y_4}{\textit{$y_4\;$}}

\cnode(120.5,41){1}{x_6}
\nput[labelsep=1]{90}{x_6}{\textit{$\;x_6$}}

% 2 invisible nodes to make the box B_1
\pnode(117.5,45){B_1a}
\pnode(117.5,33){B_1b}
\ncbox[nodesep=.25cm,boxsize=7,linewidth=0.1mm,
linestyle=dashed]{B_1a}{B_1b}

% Label ``B_1''
\nput[labelsep=7]{90}{B_1a}{$B_1$}
%====================================================
% B_2
\cnode(114.5,13){1}{y_5}
\nput[labelsep=1]{270}{y_5}{\textit{$y_5$}}

\cnode(120.5,13){1}{x_7}
\nput[labelsep=1]{90}{x_7}{\textit{$x_7$}}

% 2 invisible nodes to make the box B_2
\pnode(117.5,17){B_2a}
\pnode(117.5,5){B_2b}
\ncbox[nodesep=.25cm,boxsize=7,linewidth=0.1mm,
linestyle=dashed]{B_2a}{B_2b}

% Label ``B_2''
\nput[labelsep=7]{270}{B_2b}{$B_2$}
\fi

%============================================================
% B_1
\cnode(114.5,55){1}{y_4}
\nput[labelsep=1]{90}{y_4}{\textit{$y_4\;$}}

\cnode(120.5,55){1}{x_6}
\nput[labelsep=1]{90}{x_6}{\textit{$\;x_6$}}

% 2 invisible nodes to make the box B_1
\pnode(117.5,61){B_1a}
\pnode(117.5,49){B_1b}
\ncbox[nodesep=.25cm,boxsize=7,linewidth=0.1mm,
linestyle=dashed]{B_1a}{B_1b}

% Label ``B_1''
\nput[labelsep=7]{90}{B_1a}{$B_1$}
%====================================================
% B_2
\cnode(114.5,27){1}{y_5}
\nput[labelsep=1]{270}{y_5}{\textit{$y_5$}}

\cnode(120.5,27){1}{x_7}
\nput[labelsep=1]{270}{x_7}{\textit{$x_7$}}

% 2 invisible nodes to make the box B_2
\pnode(117.5,31){B_2a}
\pnode(117.5,19){B_2b}
\ncbox[nodesep=.25cm,boxsize=7,linewidth=0.1mm,
linestyle=dashed]{B_2a}{B_2b}

% Label ``B_2''
\nput[labelsep=7]{270}{B_2b}{$B_2$}
%=====================================================
% Destination
\pnode(150,45){i_4}
\cnode(150,41){1}{y_6}
\nput[labelsep=1]{90}{y_6}{\textit{$y_6$}}
\pnode(150,39){i_5}
\cnode(150,37){1}{y_7}
\nput[labelsep=1]{270}{y_7}{\textit{$y_7$}}
\pnode(150,33){i_6}

% Box at the Destination
\ncbox[nodesep=.25cm,boxsize=6,linewidth=0.1mm,
linestyle=dashed]{i_4}{i_6}

% Label ``Destination''
\nput[labelsep=11]{0}{i_5}{\textit{D}}

% Connection between the Source and A_1
\ncline{->}{x_1}{y_1}
% Connection between the Source and A_2
\ncline[linewidth=0.7mm]{->}{x_1}{y_2}
% Connection between the Source and A_3
\ncline[linewidth=0.7mm,linecolor=blue,linestyle=dashed]{->}{x_2}{y_3}

% Connection from A_1
\ncline{->}{x_3}{y_4}

% Connection from A_2
\ncline[linewidth=0.7mm]{->}{x_4}{y_4}
\ncline{->}{x_4}{y_5}

% Connection from A_3
\ncline{->}{x_5}{y_4}
\ncline{->}{x_5}{y_5}

% Connection between B_1 and the Destination
\ncline[linewidth=0.7mm]{->}{x_6}{y_6}

% Connection between B_2 and the Destination
\ncline{->}{x_7}{y_7}

\end{pspicture}
\caption{\small Path $\mathcal{P}_1$ identified during the first iteration is depicted in bold. During the second iteration, path $\mathcal{P}_2$ reached node $A_3$. }
\label{ex5}
%\end{scriptsize}
\end{center}
\end{figure}

\begin{example}{\em Example of rewiring across layers.}
\label{example_4}
%=================================================================================
Consider the network depicted in Fig.~\ref{ex5} and assume that the first iteration identified the path $\mathcal{P}_1=\{(x_2,y_2), \; (x_3,y_6), x_5,y_9)\}$.
During the second iteration, path $\mathcal{P}_2$ reaches and marks node $A_3$,
as depicted in  Fig.~\ref{ex5}. Assume that the algorithm then explores the edge $(x_5,y_4)$
and performs the $L_x$ function.
In this case we have that
\[
\mathbf{T}( \{x_4,x_5\},\{y_4\})=
\begin{array}{cc}
& \begin{array}{c}  y_4  \\
  \end{array}\\
\begin{array}{c}  x_4   \\ x_5\\
  \end{array}
&  \left( \begin{array}{c}
  1  \\
  1  
\end{array} \right)\end{array}, \quad \mbox{ and } L_{x_5}=\{x_4\}. 
\]
We thus visit node $A_2=A(x_4)$. Since it is the first time we visit this node, we perform the $\phi$-function at node $A_2$.
That is, at the first layer, where we now have
\[
U=\{(x_1,y_2), (x_2,y_3)\}, \quad U_x=\{x_1,x_2\},  
 U_y=\{y_2,y_3\},\quad \mathbf{T}(U_x,U_y)=
\begin{array}{cc}
& \begin{array}{cc}  y_2 & y_3  \\
  \end{array}\\
\begin{array}{c}  x_1   \\ x_2\\
  \end{array}
&  \left( \begin{array}{cc}
 1 & 1  \\
 0 & 1  
\end{array} \right)\end{array}
\]
we no longer need to use the output $y_2$, and thus can explore inputs $x_1$ and $x_2$. From $x_2$ we cannot proceed.
From $x_1$ we can use the edge $(x_1,y_1)$ and reach node $A_1$ as depicted in Fig.~\ref{ex5_1}. We do not need perform the $\phi$ function at $A_1$ as there is no additional path using this node. We proceed to explore the edge $(x_3,y_4)$ 
 and perform the $L_x$ function for $x_3$.

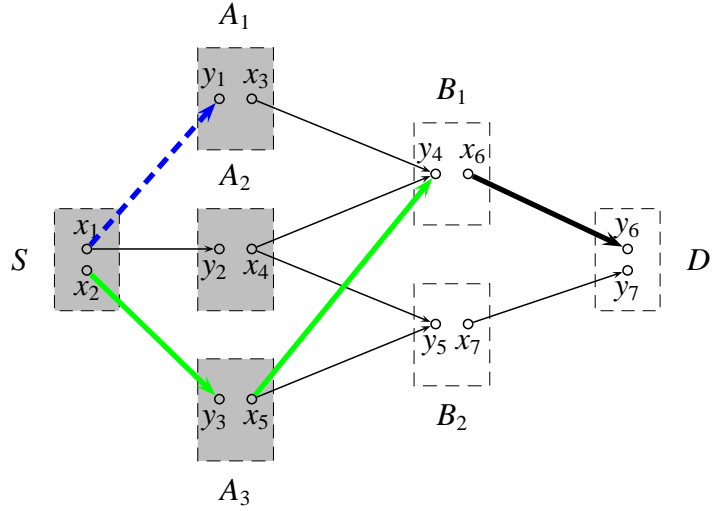
\begin{figure}[!t]
\begin{center}
%\begin{scriptsize}
\psset{unit=0.028in}
\begin{pspicture}(30,-10)(175,80)
\psset{linewidth=0.2mm}

% Source

% Box at the Source
\pnode(50,45){i_1}
\pnode(50,33){i_3}
\ncbox[fillcolor=lightgray, fillstyle=solid,nodesep=.25cm,boxsize=6,linewidth=0.1mm, linestyle=dashed]{i_1}{i_3}

\cnode(50,41){1}{x_1}
\nput[labelsep=1]{90}{x_1}{\textit{$x_1$}}

\cnode(50,37){1}{x_2}
\nput[labelsep=1]{270}{x_2}{\textit{$x_2$}}

% Label ``Source''
\pnode(50,39){i_2}
\nput[labelsep=11]{180}{i_2}{\textit{S}}
%===============================================

%  node A_1
% 2 invisible nodes to make the box A_1
\pnode(77.5,75){A_1a}
\pnode(77.5,63){A_1b}
\ncbox[fillcolor=lightgray, fillstyle=solid,nodesep=.25cm,boxsize=7,linewidth=0.1mm,
linestyle=dashed]{A_1a}{A_1b}

\cnode(74.5,69){1}{y_1}
\nput[labelsep=1]{90}{y_1}{\textit{$y_1\;$}}
%\cnode(74.5,62){1}{y_2}
%\nput[labelsep=1]{270}{y_2}{\textit{$y_2\;$}}

\cnode(80.5,69){1}{x_3}
\nput[labelsep=1]{90}{x_3}{\textit{$\;x_3$}}

% Label ``A_1''
\nput[labelsep=7]{90}{A_1a}{$A_1$}

%===========================================

% New node A_2
% 2 invisible nodes to make the box A_2
\pnode(77.5,45){A_2a}
\pnode(77.5,33){A_2b}
\ncbox[fillcolor=lightgray, fillstyle=solid,nodesep=.25cm,boxsize=7,linewidth=0.1mm, linestyle=dashed]{A_2a}{A_2b}

\cnode(74.5,41){1}{y_2}
\nput[labelsep=1]{270}{y_2}{\textit{$y_2\;$}}
%\cnode(80,42){1}{y_2}
%\nput[labelsep=1]{270}{y_2}{\textit{$y_4\;$}}

\cnode(80.5,41){1}{x_4}
\nput[labelsep=1]{270}{x_4}{\textit{$\;x_4$}}

% Label ``A_2''
\nput[labelsep=7]{90}{A_2a}{$A_2$}

%===============================
%=============================
% A_3
% 2 invisible nodes to make the box A_4
\pnode(77.5,17){A_3a}
\pnode(77.5,5){A_3b}
\ncbox[fillcolor=lightgray, fillstyle=solid,nodesep=.25cm,boxsize=7,linewidth=0.1mm,
linestyle=dashed]{A_3a}{A_3b}

\cnode(74.5,13){1}{y_3}
\nput[labelsep=1]{270}{y_3}{\textit{$y_3\;$}}

\cnode(80.5,13){1}{x_5}
\nput[labelsep=1]{270}{x_5}{\textit{$\;x_5$}}

% Label ``A_2''
\nput[labelsep=7]{270}{A_3b}{$A_3$}

%==================================================
\iffalse
% B_1
\cnode(114.5,41){1}{y_4}
\nput[labelsep=1]{90}{y_4}{\textit{$y_4\;$}}

\cnode(120.5,41){1}{x_6}
\nput[labelsep=1]{90}{x_6}{\textit{$\;x_6$}}

% 2 invisible nodes to make the box B_1
\pnode(117.5,45){B_1a}
\pnode(117.5,33){B_1b}
\ncbox[nodesep=.25cm,boxsize=7,linewidth=0.1mm,
linestyle=dashed]{B_1a}{B_1b}

% Label ``B_1''
\nput[labelsep=7]{90}{B_1a}{$B_1$}
%====================================================
% B_2
\cnode(114.5,13){1}{y_5}
\nput[labelsep=1]{270}{y_5}{\textit{$y_5$}}

\cnode(120.5,13){1}{x_7}
\nput[labelsep=1]{90}{x_7}{\textit{$x_7$}}

% 2 invisible nodes to make the box B_2
\pnode(117.5,17){B_2a}
\pnode(117.5,5){B_2b}
\ncbox[nodesep=.25cm,boxsize=7,linewidth=0.1mm,
linestyle=dashed]{B_2a}{B_2b}

% Label ``B_2''
\nput[labelsep=7]{270}{B_2b}{$B_2$}
\fi

%============================================================
% B_1
\cnode(114.5,55){1}{y_4}
\nput[labelsep=1]{90}{y_4}{\textit{$y_4\;$}}

\cnode(120.5,55){1}{x_6}
\nput[labelsep=1]{90}{x_6}{\textit{$\;x_6$}}

% 2 invisible nodes to make the box B_1
\pnode(117.5,61){B_1a}
\pnode(117.5,49){B_1b}
\ncbox[nodesep=.25cm,boxsize=7,linewidth=0.1mm,
linestyle=dashed]{B_1a}{B_1b}

% Label ``B_1''
\nput[labelsep=7]{90}{B_1a}{$B_1$}
%====================================================
% B_2
\cnode(114.5,27){1}{y_5}
\nput[labelsep=1]{270}{y_5}{\textit{$y_5$}}

\cnode(120.5,27){1}{x_7}
\nput[labelsep=1]{270}{x_7}{\textit{$x_7$}}

% 2 invisible nodes to make the box B_2
\pnode(117.5,31){B_2a}
\pnode(117.5,19){B_2b}
\ncbox[nodesep=.25cm,boxsize=7,linewidth=0.1mm,
linestyle=dashed]{B_2a}{B_2b}

% Label ``B_2''
\nput[labelsep=7]{270}{B_2b}{$B_2$}
%=====================================================
% Destination
\pnode(150,45){i_4}
\cnode(150,41){1}{y_6}
\nput[labelsep=1]{90}{y_6}{\textit{$y_6$}}
\pnode(150,39){i_5}
\cnode(150,37){1}{y_7}
\nput[labelsep=1]{270}{y_7}{\textit{$y_7$}}
\pnode(150,33){i_6}

% Box at the Destination
\ncbox[nodesep=.25cm,boxsize=6,linewidth=0.1mm,
linestyle=dashed]{i_4}{i_6}

% Label ``Destination''
\nput[labelsep=11]{0}{i_5}{\textit{D}}

% Connection between the Source and A_1
\ncline[linewidth=0.7mm,linecolor=blue,linestyle=dashed]{->}{x_1}{y_1}
% Connection between the Source and A_2
\ncline{->}{x_1}{y_2}
% Connection between the Source and A_3
\ncline[linewidth=0.7mm,linecolor=green]{->}{x_2}{y_3}

% Connection from A_1
\ncline{->}{x_3}{y_4}

% Connection from A_2
\ncline{->}{x_4}{y_4}
\ncline{->}{x_4}{y_5}

% Connection from A_3
\ncline[linewidth=0.7mm,linecolor=green]{->}{x_5}{y_4}
\ncline{->}{x_5}{y_5}

% Connection between B_1 and the Destination
\ncline[linewidth=0.7mm]{->}{x_6}{y_6}

% Connection between B_2 and the Destination
\ncline{->}{x_7}{y_7}

\end{pspicture}
\caption{\small Continuing from Fig. \ref{ex5}. Resulting configuration after performing the $L_x$ function for edge $(x_5,y_4)$ and the $\phi$-function at node $A_2$. The potential path $\mathcal{P}_2$ now reaches node $A_1$.}
\label{ex5_1}
%\end{scriptsize}
\end{center}
\end{figure}
Given that
\[
\mathbf{T}( \{x_3,x_5\},\{y_4\})=
\begin{array}{cc}
& \begin{array}{c}  y_4  \\
  \end{array}\\
\begin{array}{c}  x_3   \\ x_5\\
  \end{array}
&  \left( \begin{array}{c}
  1  \\
  1  
\end{array} \right)\end{array}, \quad \mbox{ and } L_{x_5}=\{x_3\}, 
\]
\textcolor{black}{we proceed to re-examine $x_5$}.
Because
\[
\mathbf{T}( \{x_3,x_5\},\{y_4,y_5\})=
\begin{array}{cc}
& \begin{array}{cc}  y_4 & y_5  \\
  \end{array}\\
\begin{array}{c}  x_3  \\ x_5\\
  \end{array}
&  \left( \begin{array}{cc}
 1 & 0  \\
 1 & 1  
\end{array} \right)\end{array}\]
we can now use this edge and proceed to node $B_2$. From node $B_2$ we can use edge $(x_7,y_7)$ to reach the destination and complete path $\mathcal{P}_2$.

Note that this is the second time during this iteration that we examine edge $(x_5,y_5)$. The first time we could not use this edge, due to LD with the used edge $(x_4,y_4)$. However, after the rewiring, the used edge in this layer became instead $(x_3,y_4)$, which is LI from $(x_5,y_5)$.

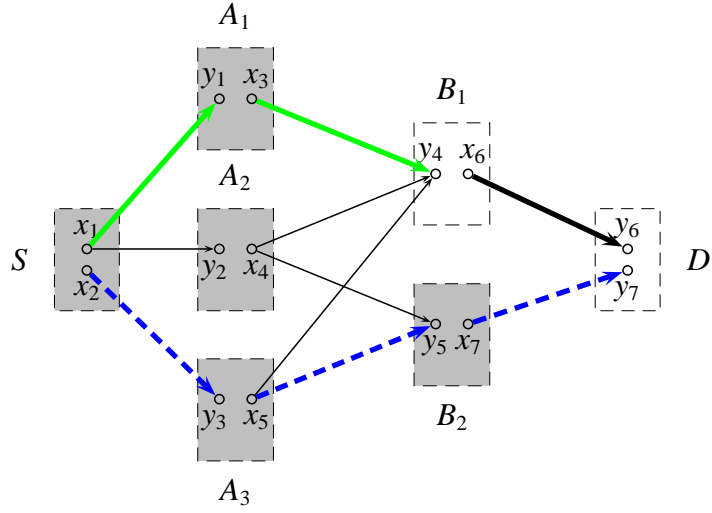
\begin{figure}[!t]
\begin{center}
%\begin{scriptsize}
\psset{unit=0.028in}
\begin{pspicture}(30,-10)(175,80)
\psset{linewidth=0.2mm}

% Source

% Box at the Source
\pnode(50,45){i_1}
\pnode(50,33){i_3}
\ncbox[fillcolor=lightgray, fillstyle=solid,nodesep=.25cm,boxsize=6,linewidth=0.1mm, linestyle=dashed]{i_1}{i_3}

\cnode(50,41){1}{x_1}
\nput[labelsep=1]{90}{x_1}{\textit{$x_1$}}

\cnode(50,37){1}{x_2}
\nput[labelsep=1]{270}{x_2}{\textit{$x_2$}}

% Label ``Source''
\pnode(50,39){i_2}
\nput[labelsep=11]{180}{i_2}{\textit{S}}
%===============================================

%  node A_1
% 2 invisible nodes to make the box A_1
\pnode(77.5,75){A_1a}
\pnode(77.5,63){A_1b}
\ncbox[fillcolor=lightgray, fillstyle=solid,nodesep=.25cm,boxsize=7,linewidth=0.1mm,
linestyle=dashed]{A_1a}{A_1b}

\cnode(74.5,69){1}{y_1}
\nput[labelsep=1]{90}{y_1}{\textit{$y_1\;$}}
%\cnode(74.5,62){1}{y_2}
%\nput[labelsep=1]{270}{y_2}{\textit{$y_2\;$}}

\cnode(80.5,69){1}{x_3}
\nput[labelsep=1]{90}{x_3}{\textit{$\;x_3$}}

% Label ``A_1''
\nput[labelsep=7]{90}{A_1a}{$A_1$}

%===========================================

% New node A_2
% 2 invisible nodes to make the box A_2
\pnode(77.5,45){A_2a}
\pnode(77.5,33){A_2b}
\ncbox[fillcolor=lightgray, fillstyle=solid,nodesep=.25cm,boxsize=7,linewidth=0.1mm, linestyle=dashed]{A_2a}{A_2b}

\cnode(74.5,41){1}{y_2}
\nput[labelsep=1]{270}{y_2}{\textit{$y_2\;$}}
%\cnode(80,42){1}{y_2}
%\nput[labelsep=1]{270}{y_2}{\textit{$y_4\;$}}

\cnode(80.5,41){1}{x_4}
\nput[labelsep=1]{270}{x_4}{\textit{$\;x_4$}}

% Label ``A_2''
\nput[labelsep=7]{90}{A_2a}{$A_2$}

%===============================
%=============================
% A_3
% 2 invisible nodes to make the box A_4
\pnode(77.5,17){A_3a}
\pnode(77.5,5){A_3b}
\ncbox[fillcolor=lightgray, fillstyle=solid,nodesep=.25cm,boxsize=7,linewidth=0.1mm,
linestyle=dashed]{A_3a}{A_3b}

\cnode(74.5,13){1}{y_3}
\nput[labelsep=1]{270}{y_3}{\textit{$y_3\;$}}

\cnode(80.5,13){1}{x_5}
\nput[labelsep=1]{270}{x_5}{\textit{$\;x_5$}}

% Label ``A_2''
\nput[labelsep=7]{270}{A_3b}{$A_3$}

%==================================================
\iffalse
% B_1
\cnode(114.5,41){1}{y_4}
\nput[labelsep=1]{90}{y_4}{\textit{$y_4\;$}}

\cnode(120.5,41){1}{x_6}
\nput[labelsep=1]{90}{x_6}{\textit{$\;x_6$}}

% 2 invisible nodes to make the box B_1
\pnode(117.5,45){B_1a}
\pnode(117.5,33){B_1b}
\ncbox[nodesep=.25cm,boxsize=7,linewidth=0.1mm,
linestyle=dashed]{B_1a}{B_1b}

% Label ``B_1''
\nput[labelsep=7]{90}{B_1a}{$B_1$}
%====================================================
% B_2
\cnode(114.5,13){1}{y_5}
\nput[labelsep=1]{270}{y_5}{\textit{$y_5$}}

\cnode(120.5,13){1}{x_7}
\nput[labelsep=1]{90}{x_7}{\textit{$x_7$}}

% 2 invisible nodes to make the box B_2
\pnode(117.5,17){B_2a}
\pnode(117.5,5){B_2b}
\ncbox[nodesep=.25cm,boxsize=7,linewidth=0.1mm,
linestyle=dashed]{B_2a}{B_2b}

% Label ``B_2''
\nput[labelsep=7]{270}{B_2b}{$B_2$}
\fi

%============================================================
% B_1
\cnode(114.5,55){1}{y_4}
\nput[labelsep=1]{90}{y_4}{\textit{$y_4\;$}}

\cnode(120.5,55){1}{x_6}
\nput[labelsep=1]{90}{x_6}{\textit{$\;x_6$}}

% 2 invisible nodes to make the box B_1
\pnode(117.5,61){B_1a}
\pnode(117.5,49){B_1b}
\ncbox[nodesep=.25cm,boxsize=7,linewidth=0.1mm,
linestyle=dashed]{B_1a}{B_1b}

% Label ``B_1''
\nput[labelsep=7]{90}{B_1a}{$B_1$}
%====================================================
% B_2
% 2 invisible nodes to make the box B_2
\pnode(117.5,31){B_2a}
\pnode(117.5,19){B_2b}
\ncbox[fillcolor=lightgray, fillstyle=solid,nodesep=.25cm,boxsize=7,linewidth=0.1mm,linestyle=dashed]{B_2a}{B_2b}

\cnode(114.5,27){1}{y_5}
\nput[labelsep=1]{270}{y_5}{\textit{$y_5$}}

\cnode(120.5,27){1}{x_7}
\nput[labelsep=1]{270}{x_7}{\textit{$x_7$}}

% Label ``B_2''
\nput[labelsep=7]{270}{B_2b}{$B_2$}
%=====================================================
% Destination
\pnode(150,45){i_4}
\cnode(150,41){1}{y_6}
\nput[labelsep=1]{90}{y_6}{\textit{$y_6$}}
\pnode(150,39){i_5}
\cnode(150,37){1}{y_7}
\nput[labelsep=1]{270}{y_7}{\textit{$y_7$}}
\pnode(150,33){i_6}

% Box at the Destination
\ncbox[nodesep=.25cm,boxsize=6,linewidth=0.1mm,
linestyle=dashed]{i_4}{i_6}

% Label ``Destination''
\nput[labelsep=11]{0}{i_5}{\textit{D}}

% Connection between the Source and A_1
\ncline[linewidth=0.7mm,linecolor=green]{->}{x_1}{y_1}
% Connection between the Source and A_2
\ncline{->}{x_1}{y_2}
% Connection between the Source and A_3
\ncline[linewidth=0.7mm,linecolor=blue,linestyle=dashed]{->}{x_2}{y_3}

% Connection from A_1
\ncline[linewidth=0.7mm,linecolor=green]{->}{x_3}{y_4}

% Connection from A_2
\ncline{->}{x_4}{y_4}
\ncline{->}{x_4}{y_5}

% Connection from A_3
\ncline{->}{x_5}{y_4}
\ncline[linewidth=0.7mm,linecolor=blue,linestyle=dashed]{->}{x_5}{y_5}

% Connection between B_1 and the Destination
\ncline[linewidth=0.7mm]{->}{x_6}{y_6}

% Connection between B_2 and the Destination
\ncline[linewidth=0.7mm,linecolor=blue,linestyle=dashed]{->}{x_7}{y_7}

\end{pspicture}
\caption{\small  Continuing from Fig. \ref{ex5_1}.  Resulting configuration after performing the $L_x$ function for edge $(x_3,y_4)$, and continuing $\mathcal{P}_2$ from node $A_3$ to node $B_2$ and $D$. }
\label{ex5_2}
%\end{scriptsize}
\end{center}
\end{figure}
\hfill{$\square$}
\end{example}

\begin{example}{\em Operations over a non-binary field.}
Consider the network depicted in Fig.~\ref{ex4}, which is similar to the network in Fig. \ref{fig_net1}, only now there is a fixed coefficient
associated with each edge over $\mathbf{F}_4$. We assume that all these coefficients equal $1$, apart from the coefficient associated with the edge $(x_4,y_7)$ that equals $2$. Operations are now over the field $\mathbf{F}_4$. For example, $y_7=2x_4+x_3$.

\begin{figure}[!t]
\begin{center}
%\begin{scriptsize}
\psset{unit=0.028in}
\begin{pspicture}(30,-10)(175,80)
\psset{linewidth=0.2mm}

% Source
\pnode(50,45){i_1}
\cnode(50,41){1}{x_1}
\nput[labelsep=1]{90}{x_1}{\textit{$x_1$}}
\pnode(50,39){i_2}
\cnode(50,37){1}{x_2}
\nput[labelsep=1]{270}{x_2}{\textit{$x_2$}}
\pnode(50,33){i_3}

% Box at the Source
\ncbox[nodesep=.25cm,boxsize=6,linewidth=0.1mm,
linestyle=dashed]{i_1}{i_3}

% Label ``Source''
\nput[labelsep=11]{180}{i_2}{\textit{S}}

% A_1
\cnode(74.5,66){1}{y_1}
\nput[labelsep=1]{90}{y_1}{\textit{$y_1\;$}}
\cnode(74.5,62){1}{y_2}
\nput[labelsep=1]{270}{y_2}{\textit{$y_2\;$}}

\cnode(80.5,66){1}{x_3}
\nput[labelsep=1]{90}{x_3}{\textit{$\;x_3$}}

% 2 invisible nodes to make the box A_1
\pnode(77.5,70){A_1a}
\pnode(77.5,58){A_1b}
\ncbox[nodesep=.25cm,boxsize=7,linewidth=0.1mm,
linestyle=dashed]{A_1a}{A_1b}

% Label ``A_1''
\nput[labelsep=7]{90}{A_1a}{$A_1$}

% A_2
\cnode(74.5,13){1}{y_3}
\nput[labelsep=1]{270}{y_3}{\textit{$y_3\;$}}

\cnode(80.5,13){1}{x_4}
\nput[labelsep=1]{270}{x_4}{\textit{$\;x_4$}}

% 2 invisible nodes to make the box A_2
\pnode(77.5,17){A_2a}
\pnode(77.5,5){A_2b}
\ncbox[nodesep=.25cm,boxsize=7,linewidth=0.1mm,
linestyle=dashed]{A_2a}{A_2b}

% Label ``A_2''
\nput[labelsep=7]{270}{A_2b}{$A_2$}

% B_1
\cnode(114.5,66){1}{y_6}
\nput[labelsep=1]{90}{y_6}{\textit{$y_6\;$}}

\cnode(120.5,66){1}{x_5}
\nput[labelsep=1]{90}{x_5}{\textit{$\;x_5$}}

% 2 invisible nodes to make the box B_1
\pnode(117.5,70){B_1a}
\pnode(117.5,58){B_1b}
\ncbox[nodesep=.25cm,boxsize=7,linewidth=0.1mm,
linestyle=dashed]{B_1a}{B_1b}

% Label ``B_1''
\nput[labelsep=7]{90}{B_1a}{$B_1$}

% B_2
\cnode(114.5,13){1}{y_7}
\nput[labelsep=1]{270}{y_7}{\textit{$y_7$}}

\cnode(120.5,13){1}{x_6}
\nput[labelsep=1]{90}{x_6}{\textit{$x_6$}}
\cnode(120.5,9){1}{x_7}
\nput[labelsep=1]{270}{x_7}{\textit{$x_7$}}

% 2 invisible nodes to make the box B_2
\pnode(117.5,17){B_2a}
\pnode(117.5,5){B_2b}
\ncbox[nodesep=.25cm,boxsize=7,linewidth=0.1mm,
linestyle=dashed]{B_2a}{B_2b}

% Label ``B_2''
\nput[labelsep=7]{270}{B_2b}{$B_2$}

% Destination
\pnode(150,45){i_4}
\cnode(150,41){1}{y_8}
\nput[labelsep=1]{90}{y_8}{\textit{$y_8$}}
\pnode(150,39){i_5}
\cnode(150,37){1}{y_9}
\nput[labelsep=1]{270}{y_9}{\textit{$y_9$}}
\pnode(150,33){i_6}

% Box at the Destination
\ncbox[nodesep=.25cm,boxsize=6,linewidth=0.1mm,
linestyle=dashed]{i_4}{i_6}

% Label ``Destination''
\nput[labelsep=11]{0}{i_5}{\textit{D}}

% Connection between the Source and A_1
\ncline{->}{x_1}{y_1}
\ncline[linewidth=0.7mm]{->}{x_2}{y_2}

% Connection between the Source and A_2
\ncline[linewidth=0.7mm,linecolor=blue,linestyle=dashed]{->}{x_1}{y_3}

% Connection from A_1
\ncline[linewidth=0.7mm]{->}{x_3}{y_6}\Aput{1}
\ncline{->}{x_3}{y_7}\Aput{1}

% Connection from A_2
\ncline{->}{x_4}{y_6}\Aput{1}
\ncline[linewidth=0.7mm,linecolor=blue,linestyle=dashed]{->}{x_4}{y_7}\Aput{2}

% Connection between B_1 and the Destination
\ncline[linewidth=0.7mm]{->}{x_5}{y_9}

% Connection between B_2 and the Destination
\ncline[linewidth=0.7mm,linecolor=blue,linestyle=dashed]{->}{x_6}{y_8}
\ncline{->}{x_7}{y_9}

\end{pspicture}
\caption{\small An example of a  nonbinary linear deterministic network. Each edge is associated with a coefficient over  $\mathbf{F}_4$.  All these coefficients equal $1$, apart from the coefficient associated with the edge $(x_4,y_7)$ that equals $2$. }
\label{ex4}
%\end{scriptsize}
\end{center}
\end{figure}
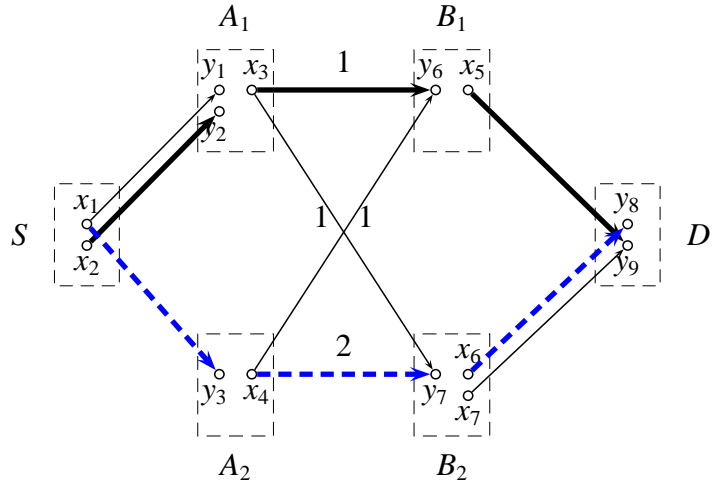

Assume that the first iteration identified the path $\mathcal{P}_1=\{(x_2,y_2), \; (x_3,y_6), (x_5,y_9) \}$. 
During the second iteration, assume that we use at the first layer the edge  $(x_1,y_3)$, and arrive at layer $2$.  
At this layer, $U_x=\{x_3\}$ and $U_y=\{y_6\}$. To use edge $(x_4,y_7)$, we examine whether the matrix
\[
\mathbf{T}(\{x_3,x_4\},\{y_6,y_7\})=
\begin{array}{cc}
& \begin{array}{cc}  y_6 & y_7  \\
  \end{array}\\
\begin{array}{c}  x_3  \\ x_4\\
  \end{array}
&  \left( \begin{array}{cc}
 1 & 1  \\
 1 & 2  
\end{array} \right)\end{array}
\]
 is full rank over $\mathbf{F}_4$. As indeed it is, we can reach node $B_2$, and from there using edge $(x_6,y_8)$ complete $\mathcal{P}_2$.
Note that in the binary example in Fig.~\ref{fig_net1}, we could only identify one path.
\hfill{$\square$}
\end{example}

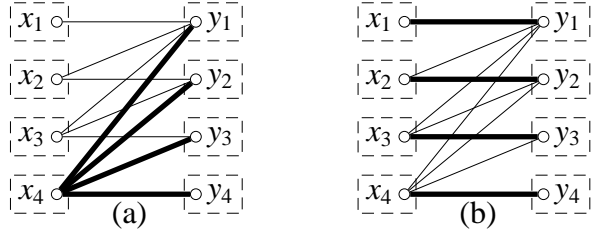
\begin{figure}[!t]
\begin{center}
\psset{unit=0.03in}
\begin{pspicture}(40,18)(135,50)
\psset{linewidth=0.1mm}
%%% First Figure
% Vertices
\cnode(46,50){1}{x_1}
\nput[labelsep=1]{180}{x_1}{\textit{$x_1$}}
\pnode(43,52){x_1a}
\pnode(43,48){x_1b}
\ncbox[nodesep=.1cm,boxsize=5,linewidth=0.1mm,
linestyle=dashed]{x_1a}{x_1b}

\cnode(46,40){1}{x_2}
\nput[labelsep=1]{180}{x_2}{\textit{$x_2$}}
\pnode(43,42){x_2a}
\pnode(43,38){x_2b}
\ncbox[nodesep=.1cm,boxsize=5,linewidth=0.1mm,
linestyle=dashed]{x_2a}{x_2b}

\cnode(46,30){1}{x_3}
\nput[labelsep=1]{180}{x_3}{\textit{$x_3$}}
\pnode(43,32){x_3a}
\pnode(43,28){x_3b}
\ncbox[nodesep=.1cm,boxsize=5,linewidth=0.1mm,
linestyle=dashed]{x_3a}{x_3b}

\cnode(46,20){1}{x_4}
\nput[labelsep=1]{180}{x_4}{\textit{$x_4$}}
\pnode(43,22){x_4a}
\pnode(43,18){x_4b}
\ncbox[nodesep=.1cm,boxsize=5,linewidth=0.1mm,
linestyle=dashed]{x_4a}{x_4b}

\cnode(70,50){1}{y_1}
\nput[labelsep=1]{0}{y_1}{\textit{$y_1$}}
\pnode(73,52){y_1a}
\pnode(73,48){y_1b}
\ncbox[nodesep=.1cm,boxsize=5,linewidth=0.1mm,
linestyle=dashed]{y_1a}{y_1b}

\cnode(70,40){1}{y_2}
\nput[labelsep=1]{0}{y_2}{\textit{$y_2$}}
\pnode(73,42){y_2a}
\pnode(73,38){y_2b}
\ncbox[nodesep=.1cm,boxsize=5,linewidth=0.1mm,
linestyle=dashed]{y_2a}{y_2b}

\cnode(70,30){1}{y_3}
\nput[labelsep=1]{0}{y_3}{\textit{$y_3$}}
\pnode(73,32){y_3a}
\pnode(73,28){y_3b}
\ncbox[nodesep=.1cm,boxsize=5,linewidth=0.1mm,
linestyle=dashed]{y_3a}{y_3b}

\cnode(70,20){1}{y_4}
\nput[labelsep=1]{0}{y_4}{\textit{$y_4$}}
\pnode(73,22){y_4a}
\pnode(73,18){y_4b}
\ncbox[nodesep=.1cm,boxsize=5,linewidth=0.1mm,
linestyle=dashed]{y_4a}{y_4b}

\put(55.5,15){(a)}
% Edges
\ncline{-}{x_1}{y_1}
%\ncline{-}{x_1}{y_2}
%\ncline{-}{x_1}{y_3}
%\ncline{-}{x_1}{y_4}
%\ncline{-}{x_1}{y_5}
%\ncline{-}{x_1}{y_6}
\ncline{-}{x_2}{y_1}
\ncline{-}{x_2}{y_2}
%\ncline{-}{x_2}{y_3}
%\ncline{-}{x_2}{y_4}
%\ncline{-}{x_2}{y_5}
%\ncline{-}{x_2}{y_6}
\ncline{-}{x_3}{y_1}
\ncline{-}{x_3}{y_2}
\ncline{-}{x_3}{y_3}
%\ncline{-}{x_3}{y_4}
%\ncline{-}{x_3}{y_5}
%\ncline{-}{x_3}{y_6}
\ncline[linewidth=0.7mm]{-}{x_4}{y_1}
\ncline[linewidth=0.7mm]{-}{x_4}{y_2}
\ncline[linewidth=0.7mm]{-}{x_4}{y_3}
\ncline[linewidth=0.7mm]{-}{x_4}{y_4}
%\ncline{-}{x_4}{y_5}
%\ncline{-}{x_4}{y_6}
\ncline{-}{x_5}{y_1}
\ncline{-}{x_5}{y_2}
\ncline{-}{x_5}{y_3}
\ncline{-}{x_5}{y_4}
\ncline{-}{x_5}{y_5}
%\ncline{-}{x_5}{y_6}
\ncline{-}{x_6}{y_1}
\ncline{-}{x_6}{y_2}
\ncline{-}{x_6}{y_3}
\ncline{-}{x_6}{y_4}
\ncline{-}{x_6}{y_5}
\ncline{-}{x_6}{y_6}

%%%%%%%% Second Figure

\cnode(106,50){1}{x_1}
\nput[labelsep=1]{180}{x_1}{\textit{$x_1$}}
\pnode(103,52){x_1a}
\pnode(103,48){x_1b}
\ncbox[nodesep=.1cm,boxsize=5,linewidth=0.1mm,
linestyle=dashed]{x_1a}{x_1b}

\cnode(106,40){1}{x_2}
\nput[labelsep=1]{180}{x_2}{\textit{$x_2$}}
\pnode(103,42){x_2a}
\pnode(103,38){x_2b}
\ncbox[nodesep=.1cm,boxsize=5,linewidth=0.1mm,
linestyle=dashed]{x_2a}{x_2b}

\cnode(106,30){1}{x_3}
\nput[labelsep=1]{180}{x_3}{\textit{$x_3$}}
\pnode(103,32){x_3a}
\pnode(103,28){x_3b}
\ncbox[nodesep=.1cm,boxsize=5,linewidth=0.1mm,
linestyle=dashed]{x_3a}{x_3b}

\cnode(106,20){1}{x_4}
\nput[labelsep=1]{180}{x_4}{\textit{$x_4$}}
\pnode(103,22){x_4a}
\pnode(103,18){x_4b}
\ncbox[nodesep=.1cm,boxsize=5,linewidth=0.1mm,
linestyle=dashed]{x_4a}{x_4b}

\cnode(130,50){1}{y_1}
\nput[labelsep=1]{0}{y_1}{\textit{$y_1$}}
\pnode(133,52){y_1a}
\pnode(133,48){y_1b}
\ncbox[nodesep=.1cm,boxsize=5,linewidth=0.1mm,
linestyle=dashed]{y_1a}{y_1b}

\cnode(130,40){1}{y_2}
\nput[labelsep=1]{0}{y_2}{\textit{$y_2$}}
\pnode(133,42){y_2a}
\pnode(133,38){y_2b}
\ncbox[nodesep=.1cm,boxsize=5,linewidth=0.1mm,
linestyle=dashed]{y_2a}{y_2b}

\cnode(130,30){1}{y_3}
\nput[labelsep=1]{0}{y_3}{\textit{$y_3$}}
\pnode(133,32){y_3a}
\pnode(133,28){y_3b}
\ncbox[nodesep=.1cm,boxsize=5,linewidth=0.1mm,
linestyle=dashed]{y_3a}{y_3b}

\cnode(130,20){1}{y_4}
\nput[labelsep=1]{0}{y_4}{\textit{$y_4$}}
\pnode(133,22){y_4a}
\pnode(133,18){y_4b}
\ncbox[nodesep=.1cm,boxsize=5,linewidth=0.1mm,
linestyle=dashed]{y_4a}{y_4b}

\put(115.5,15){(b)}
% Edges
\ncline[linewidth=0.7mm]{-}{x_1}{y_1}
%\ncline{-}{x_1}{y_2}
%\ncline{-}{x_1}{y_3}
%\ncline{-}{x_1}{y_4}
%\ncline{-}{x_1}{y_5}
%\ncline{-}{x_1}{y_6}
\ncline{-}{x_2}{y_1}
\ncline[linewidth=0.7mm]{-}{x_2}{y_2}
%\ncline{-}{x_2}{y_3}
%\ncline{-}{x_2}{y_4}
%\ncline{-}{x_2}{y_5}
%\ncline{-}{x_2}{y_6}
\ncline{-}{x_3}{y_1}
\ncline{-}{x_3}{y_2}
\ncline[linewidth=0.7mm]{-}{x_3}{y_3}
%\ncline{-}{x_3}{y_4}
%\ncline{-}{x_3}{y_5}
%\ncline{-}{x_3}{y_6}
\ncline{-}{x_4}{y_1}
\ncline{-}{x_4}{y_2}
\ncline{-}{x_4}{y_3}
\ncline[linewidth=0.7mm]{-}{x_4}{y_4}
%\ncline{-}{x_4}{y_5}
%\ncline{-}{x_4}{y_6}
\ncline{-}{x_5}{y_1}
\ncline{-}{x_5}{y_2}
\ncline{-}{x_5}{y_3}
\ncline{-}{x_5}{y_4}
\ncline[linewidth=0.7mm]{-}{x_5}{y_5}
%\ncline{-}{x_5}{y_6}
\ncline{-}{x_6}{y_1}
\ncline{-}{x_6}{y_2}
\ncline{-}{x_6}{y_3}
\ncline{-}{x_6}{y_4}
\ncline{-}{x_6}{y_5}
\ncline[linewidth=0.7mm]{-}{x_6}{y_6}

\end{pspicture}
\caption{\small A layer-cut and (a) the traditional approach where interference is treated as noise, (b) the approach where interference 
is allowed.}
\label{int1}
\end{center}
\end{figure}

We conclude with an example that shows the benefits of not treating interference as noise. 
%\subsection{Example 5}
\begin{example}{\em Benefits from constructive use of interference.}
The traditional approach adopted today in wireless networks is that if one or more transmitted signals interfere with a received signal, they are treated as noise. Such interference is avoided through scheduling. This approach can lead to significant loss of capacity. Consider a network that has the layer-cut depicted in 
Fig.~\ref{int1}. Fig.~\ref{int1}(a) depicts the traditional solution:
treating interference as noise implies that we cannot simultaneously have two broadcast transmissions that interfere,
and thus we can have at most one broadcast transmission. Fig.~\ref{int1}(b) shows that, if interference is allowed, we can in fact 
use four LI edges through this cut (the example is easily generalized to $N$ nodes leading to $O(N)$ benefits). Indeed, the transfer matrix associated with this cut,
\[ \mathbf{T}(\{x_1,x_2,x_3,x_4 \},\{y_1,y_2,y_3,y_4 \}) =\left( \begin{array}{cccc}
1 & 1 & 1 & 1\\
0 & 1 & 1 & 1 \\
0 & 0 & 1 & 1 \\
0 & 0 & 0 & 1 \\
%0 & 0 & 0 & 0 & 1 & 1\\
%0 & 0 & 0 & 0 & 0 & 1
\end{array} \right)\]
has rank four. This matrix coincides with the transformation matrix of the highlighted edges.

%We can generalize this model to $n$ nodes transmitting and $n$ nodes receiving the information. Each $x_i$ in the bipartite graph transmits to all the $y_j$ for $j=1,\cdots,i$. We will have a capacity $C=1$ when interference is not allowed (all nodes are transmitting to $y_1$ so if more than one node is transmitting at the same time, interference will occur at $y_1$) and $C=n$ when interference is allowed (the transfer matrix of such a bipartite graph is of full rank: upper triangular matrix). In other words, the capacity remains constant when interference is allowed and it increases with $n$ when it is not.%
\hfill{$\square$}
\end{example}

%%%%%%%%%%%%%%%%%%%%%%%%%%%%%%%%%%%%%%%%%%%%%%%%%%%%%%%%%%%%%%%%%%%%%%%%%%%%%%%%%%%%%%%%

\subsection{Main Result}
Our main result is the following theorem.
\begin{theorem}\label{th1}
  The unicast algorithm identifies $\mathcal{C}$ LI paths, where $\mathcal{C}$ is the min-cut
  value between the source-destination pair in a linear deterministic network.

In particular, 
the number of the paths identified by the algorithm equals the rank of the transfer matrix between the inputs in $V_1$ and the outputs in $V_2$, where $V_1$ are the marked and $V_2$ are the unmarked nodes when the algorithm stops.
\end{theorem}

\begin{proof}
  Based on the algorithm, it is clear that when the algorithm stops, 
  the provided output is a set of linearly
  independent source-destination paths $\mathcal{P}$.   
  
   Let $K$ denote the number of these paths; this implies that the algorithm stops, i.e., fails to find an additional path, during iteration $K+1$.
  Since $K$ can never exceed the rank 
  of a source-destination cut, i.e., $K\leq \mathcal{C}$, 
  it suffices to find a cut whose capacity is not bigger than the number
  of paths identified by our algorithm. Let $V_1$ be the set of all marked
  (visited) vertices and $V_2$ be the other vertices during iteration $K+1$, when the algorithm stops.
  Clearly, $(V_1, V_2)$ is a source destination cut.  

    Consider now the matrix $\mathbf{T}(V_1, V_2)$, where, by a slight abuse of notation, the set of rows  of this  matrix correspond to the inputs $x$ in $V_1$ and the set of columns to the outputs $y$ in nodes in $V_2$ respectively. By
    appropriate ordering of these inputs and outputs we can bring the transfer matrix in to a block diagonal form, in which every block corresponds to a layer of
    the network. More precisely, if $W_i$ ($W_i'$) is the set of visited 
    (unvisited) nodes in the $i$-th layer then $\mathbf{T}(V_1, V_2)$ 
     can be regarded
     as a block diagonal matrix whose $i$-th block is $\mathbf{T} (W_i, W_{i + 1}')$.  For clarity we have collected all the notation we use in this proof in Table~\ref{table_not}.

%========================================================================================
\begin{table*}
\begin{center}\label{table_not}
\begin{tabular}{|cl|}\hline
 $\mathbf{F}_q$ & finite field of operation\\\hline
$S$ & the source node\\\hline
$D$ & the destination node\\\hline
$\Lambda$ & number of network layers \\\hline
$M$ & maximum number of nodes per layer\\ \hline
$A(x)$ & node where input $x$ belongs \\ \hline
$A(y)$ & node where output $y$ belongs \\ \hline
 $\mathbf{T} (V, W)$ & transformation matrix whose rows
  are labeled with the elements of $V$\\ 
  & and the columns with the elements of $W$ \\ \hline
$|U|$ & number of identified LI paths in previous iterations \\\hline
$U$ & set of used channels between layers $i$ and $i+1$ (we drop the index $i$ for simplicity)\\\hline
$U_x$ & set of used inputs at layer $i$ corresponding to the channels in $U$\\\hline
$U_y$ & set of used outputs at layer $i+1$ corresponding to the channels in $U$\\\hline
$\mathcal{R}^{(i)}$ &  the set of edges that input $x_i$ perceives as being used from previous iterations\\\hline
$\mathcal{R}^{(i)}_x$ & set of inputs that input $x_i$ perceives as being used from previous iterations\\\hline
$\mathcal{R}^{(i)}_y$ & set of outputs that input $x_i$ perceives as being used from previous iterations\\\hline
$L_{x_i}(Z)$ & the smallest subset of $\mathcal{R}^{(i)}$ in the matrix  $\mathbf{T}(\mathcal{R}^{(i)},Z)$ \\ 
&that contains $x_i$ in its span\\
\hline
$W_i$    & set of visited nodes in the $i$ layer \\\hline
 $W_{i}'$   & set of  unvisited nodes in the $i$-th layer \\\hline
$U_{Bx} $ & set of all the
    inputs of the nodes in $W_x$ \\\hline 
$U_{Bx}'$ & set of all the visited
    inputs in the  layer $i$ which are used, i.e. $U_{Bx'}=U_{Bx}\cap U_x$.  \\ \hline
$U_{By}$ & set of all the outputs  in $W_{i+1}'$\\ \hline
$U_{By}'$ & set of all the outputs of $U_{By}$ which are used, i.e., $U_{By'}=U_{By}\cap U_y$.\\\hline
 $I_{total}$&  total number of inputs  in the network.\\\hline
$O_{total}$ & total number of outputs in the network.\\\hline

\end{tabular}
\end{center}
\caption{ Summary of Notation}
\end{table*}
%==================================================================================================

 We will show  in Lemma~\ref{lemma_1} that for every integer $1 \leq i \leq \Lambda$ it holds that, 
\begin{equation}
\begin{aligned}\label{eq_toshow}
& \text{rank} (\mathbf{T} (W_i,
    W_{i + 1}')) \leq \\
& |\{e = vv' \in E (G) |v \in W_i, v' \in W_{i + 1}', e
    \in U\}| - |\{e = vv' \in E (G) |v \in W_i', v' \in W_{i + 1}, e \in U\}|
\end{aligned}
\end{equation}
 where recall that we denote by $U$ the set of used edges by paths in $\mathcal{P}$ at layer $i$.  
That is, if $e \in U$, then it belongs in some path  $j$, i.e., $e \in \mathcal{P}_j$
(and more generally $e$ belongs in the set of all used edges in the graph, i.e., $e \in \mathcal{P}$). 
Also, from the structure of the layer network, the total number of paths equals the number of used edges in each layer, namely, $|U|=K$.
Now
\begin{equation}  
  \begin{aligned}\label{eq1}
      &  \text{rank}  (\mathbf{T} (V_1, V_2)) \\
      &  \stackrel{(a)}{=} \sum_{i=1}^\Lambda  \text{rank} (\mathbf{T} (W_i, W_{i + 1}'))\\ 
      &   \stackrel{(b)}{\leq} |\{e = vv' |v 
        \text{ is visited but } v'  \text{ is not visited}, 
           e \in \mathcal{P} \}| -
       |\{e = vv' |v \text{ is not visited  but }\\ & v'  \text{ is visited}, 
       e \in \mathcal{P} \}| \\
      &  \stackrel{(c)}{=} K.
\end{aligned} 
   \end{equation}
    Equality $(a)$ holds from the fact that the rank of
    any block diagonal matrix is the sum of the rank of its blocks. Inequality $(b)$  follows directly from Lemma \ref{lemma_1} that  will prove in the following. Finally, equality $(c)$  holds because for each source-destination path $\mathcal{P}_i$
    the ``used'' edges by  $\mathcal{P}_i$ contribute exactly one in the
    difference, that is,
\begin{equation}  
\begin{aligned}
&|\{e = vv' |v 
        \text{ is visited but } v'  \text{ is not visited}, 
           e \in \mathcal{P}_i \}| -\\
       &- |\{e = vv' |v \text{ is not visited  but } v'  \text{ is visited}, 
       e \in \mathcal{P}_i \}|=1
\end{aligned}
\end{equation}
Indeed, given a cut $(V_1,V_2)$, with $S \in V_1$ and $D \in V_2$, for $\mathcal{P}_i$ to connect  $S$ to $D$, it 
must cross at least one time from  $V_1$ to $V_2$.
If it crosses $m \geq 1$ times from  $V_1$ to $V_2$, then it also has to cross $m-1$ times from $V_2$ to $V_1$.
%
%\textcolor{blue}{** Explain this better, perhaps with a figure, to show that 
% each path crosses the cut, and if it crosses it m times in the correct direction then it crosses it m-1 times in the reverse direction**}
   %Thus  (\ref{eq_toshow}).
   \end{proof}

%===========================================================================================++++++++++++++++++
%===========================================================================================++++++++++++++++++
\begin{lemma}\label{lemma_1}
%Assume that the algorithm stops at iteration $|U|+1$, and $W_i$, $W_i'$ are as defined in Table \ref{table_notation}.
For every integer $1 \leq i \leq \Lambda$ it holds that, 
\begin{equation}
\begin{aligned}\label{eq_toshow1}
& \text{rank} (\mathbf{T} (W_i,
    W_{i + 1}')) \leq \\
& |\{e = vv' \in E (G) |v \in W_i, v' \in W_{i + 1}', e
    \in U\}| - |\{e = vv' \in E (G) |v \in W_i', v' \in W_{i + 1}, e \in U\}|
\end{aligned}
\end{equation}
\end{lemma}
%===========================================================================================++++++++++++++++++
%===========================================================================================++++++++++++++++++

\begin{proof}     
    Fix an integer $1 \leq i \leq \Lambda$. 
      Recall that we denote by $U$ the set of used channels in this layer (dropping the index $i$ for simplicity), 
      $U_x$  their inputs and $U_y$ their outputs.    Additionally, let $U_{Bx}$ be the set of all the
    inputs of the nodes in $W_i$ and $U_{Bx}'$ be the set of all the visited
    inputs in the current layer which appear in some identified path. That is, 
$U_{Bx}'=U_{Bx}\cap U_x$.  Let 
    $U_{By}$ be the set of all the outputs  that are in the $i + 1$-st layer
    and are not visited and $U_{By}'$ be those outputs of $U_{By}$ which
    appear on some identified path (i.e., used outputs). That is, $U_{By}'=U_{By}\cap U_y$. The notation is summarized in Table~\ref{table_not}.

 We are interested in calculating the rank of the matrix  $\mathbf{T}(W_i, W_{i + 1}')=\mathbf{T} (U_{Bx}, U_{By})$. 
  Note that we can split 
    the columns of $\mathbf{T} (U_{Bx}, U_{By})$  into two parts, one corresponding to the used and unmarked outputs, i.e., $U_{By}'$, 
    and the other corresponding to the unused and unmarked outputs, $U_{By} - U_{By}'$. 
    Similarly, we can split
 the rows 
    again into two parts, one corresponding to the used and marked inputs, $U_{Bx}'$,
    and the other to the unused and marked inputs, $U_{Bx} - U_{Bx}'$.

Our proof proceeds as follows.
Lemmas~\ref{lemma_s11} and~\ref{lemma_s1}  prove that all the rows of $\mathbf{T} (U_{Bx}, U_{By})$ that belong to
    the second part (in $U_{Bx} - U_{Bx}'$ ) are in the span of the rows corresponding to the inputs in the first part (in $U_{Bx}'$). 
 As a result,
\begin{equation}
 \text{rank} \mathbf{T} (U_{Bx}, U_{By})=\text{rank}  \mathbf{T} (U_{Bx}', U_{By}).
\end{equation}
 Lemma~\ref{lemma_s3} builds on this result to
 prove that
%that removing each of these columns reduces the rank of the matrix $\mathbf{A}$ by one. 
%    Note that, if,  removing any column from a subset of columns of a
%    matrix, reduces the rank of the matrix by one, then none of the columns in that subset
%    belongs to the span of other columns. Therefore, if we remove all the
%    columns of the subset, the rank of the matrix also drops by the size of
%    the subset.
%In our case, if Lemma~\ref{lemma_s3} holds, then if we remove all the columns in the subset $U_y-U_{By}'$ from matrix $\mathbf{A}$, 
%then the resulting matrix 
 \begin{equation}\label{eq_x}
\text{rank}\mathbf{T} (U_{Bx}, U_{By})=|U_{Bx}'| - (|U_y|-|U_{By}'|).
\end{equation}
Showing that (\ref{eq_x}) holds is the main technical part of this proof.
Now we distinguish three cases for each edge $e=(x,y)\in U$:
\begin{enumerate}
\item $x \in U_{Bx}'$ and $y \notin \{U_y - U_{By}'\}$:  the edge contributes value ``one'' only in   
$|U_{Bx}'|$,
\item $x \notin U_{Bx}'$ and $y \in \{ U_y - U_{By}'\}$:  the edge contributes value ``one'' only in   $(|U_y|-|U_{By}'|)$
\item $x \in U_{Bx}'$ and $y \in \{U_y - U_{By}'\}$: then the edge contributes value ``one'' both in   
$|U_{Bx}'|$ and in $(|U_y|-|U_{By}'|)$ and thus does not affect the quantity  $|U_{Bx}'| - (|U_y|-|U_{By}'|)$.
\end{enumerate}
Thus
\begin{equation}
\begin{aligned}\nonumber
&|U_{Bx}'| - (|U_y|-|U_{By}'|)=\\
& |\{e = vv' \in E (G) |v \in U_{Bx}', v' \in U_{By}', e
    \in U\}|- |\{e = vv' \in E (G) |v \notin U_{Bx}', v' \in U_{By}, e \in U\}|\\
& =|\{e = vv' \in E (G) |v \in W_i, v' \in W_{i + 1}', e
    \in U\}| - |\{e = vv' \in E (G) |v \in W_i', v' \in W_{i + 1}, e \in U\}|
\end{aligned}
\end{equation} 
%\end{enumerate}
and our proof is concluded.
%\textcolor{blue}{Why is this quantity always positive?}
\end{proof}

%=========================================================================================
Before we %prove the  lemmas that we used as steps in the previous proof,
continue, we 
need to introduce some additional notation.

When iteration $K+1$ starts, at the layer we are examining, we have identified from the previous iterations
a set of used edges $U$, with corresponding set of inputs and outputs $U_x$ and $U_y$ respectively.
As the algorithm attempts to find the $K+1$ path, it may perform some rewirings inside this layer (due to 
consecutive executions for example of several $L_x$ and $\phi$-functions). Thus, 
when input $x_i$ gets marked and starts to be explored by the algorithm,
this input might perceive as used a different set of edges than $U$. 
We will denote by $\mathcal{R}^{(i)}$ the set of edges that input $x_i$ perceives as used (by the $K$ paths),
and $\mathcal{R}_x^{(i)}$, $\mathcal{R}_y^{(i)}$ the corresponding sets of used inputs and outputs. 
Note that, while all the edges emannating from $x_i$ are examined, for all of them the algorithm will
 assume the same set of used edges $\mathcal{R}^{(i)}$.% (for example, if no rewirings have occured, $\mathcal{R}^{(i)}=U$). 
%\textcolor{blue}{For example, in }

Now, from assumption, the iteration $K+1$ fails to find a path from $S$ to $D$. 
Thus, although several rewirings might be attempted, because iteration
$K+1$ fails, when the algorithm stops we have reverted to the original set $U$.

%==================================================
\begin{lemma}% {(\em Property 1.)} 
\label{lemma_s01}
For all $x_i \in U_{Bx}-U_{Bx}'$ it holds that
\[
\text{rank}\mathbf{T}(\{\mathcal{R}^{(i)}_{x}, x_i\}, \{ \mathcal{R}_y^{(i)}, U_{By}-U_{By}'\})=
\text{rank}\mathbf{T}(\mathcal{R}^{(i)}_{x}, \{ \mathcal{R}_y^{(i)}, U_{By}-U_{By}'\}) 
\]
In particular, there exists a minimal set of rows $L_{x_i}\subseteq \mathcal{R}^{(i)}_{x}$
such that
\begin{equation}\label{eq_L_imp}
\begin{aligned}
 \text{rank}(\mathbf{T}(\{L_{{x_i}},x_i\},\{\mathcal{R}_{y}^{(i)},U_{By}-U_{By}'\}))=
  \text{rank}(\mathbf{T}(L_{{x_i}},\{\mathcal{R}^{(i)}_{y},U_{By}-U_{By}'\})).
\end{aligned}
\end{equation}%\hfill{$\square$}
\end{lemma}

\begin{proof}
For this proof only we also use the following notation. 
%The following notation is also useful. % of minimally linearly dependent set.
Assume that  $\text{rank}(\mathbf{T}(\{\mathcal{R}^{(i)},x_i\},Z))=\text{rank}(\mathbf{T}(\mathcal{R}^{(i)},Z))$ for some set of columns $Z$.
Define  $L_{x_i}(Z)$ to be the smallest subset of $\mathcal{R}^{(i)}$ that contains $x_i$ in its span,
i.e., \[\text{rank}(\mathbf{T}(\{L_{x_i}(Z),x_i\},Z))=\text{rank}(\mathbf{T}(L_{x_i}(Z),Z)).\] 
We  will use for abbreviation $L_{x_i}=L_{x_i}(\mathcal{R}^{(i)}_y)$.

Decompose the column indices of the matrix $\mathbf{T} (\mathcal{R}^{(i)}_x, \{\mathcal{R}^{(i)}_y, U_{By}-U_{By}'\})$  in the following 4 nonoverlaping parts:  $[\mathcal{R}^{(i)}_{y},\quad W_1,\quad W_L,\quad W_0]$. Here\\
%\begin{itemize}
%\item 
$\bullet$ $\mathcal{R}^{(i)}_{y}$ contains all the used $y$'s,\\
%\item 
$\bullet$ $W_1$ contains all $y_j$ such that the edges $(x_i,y_j)$ exist but cannot be used due to LD,\\ 
%\item  
$\bullet$ $W_L$ contains all the remaining $y_k \in W_{By}$ that have at least one nonzero value in each column (i.e., the set of all $y$ columns where at least one  edge $(x_k,y_k)$ with $x_k\in L_{x_i}(W_{By})$ exists, but $x_i$ has zero value), and\\
%\item 
$\bullet$ $W_0$ contains all zero columns (this is the set of $y$'s associated with $x$'s not in the set $\{L_{{x_i}}(W_{By}),x_i\}$).\\
%\end{itemize}

We underline that the set of columns  $U_{By}-U'_{By}=\{W_1,\quad W_L,\quad W_0\}$ is the set of unmarked unused outputs {\em at the end} of the iteration $K+1$,
and is the same independently of the set of outputs in $\mathcal{R}^{(i)}_y$. 
Note that, because $\mathcal{R}^{(i)}_y$ can contain  either outputs that belong in $U_y$ (that thus are used)
and/or outputs obtained through the execution of the $\phi$-function (and thus are marked), has by definition  zero overlap with the set $U_{By}-U'_{By}$ which contains outputs that are both unmarked and not used.

To prove the lemma, it is sufficient to prove that the following equation holds.
\begin{equation}\label{prop6} 
\begin{aligned}
L_{x_i} & = L_{x_i}(\mathcal{R}^{(i)}_y)\stackrel{(a)}{=}L_{x_i}(\{\mathcal{R}^{(i)},W_1\})\\
& \stackrel{(b)}{=}L_{x_i}(\{\mathcal{R}^{(i)},W_L\})\stackrel{(c)}{=}L_{x_i}(\{\mathcal{R}^{(i)},W_0\}).
\end{aligned}
\end{equation}
{ $(a):\quad$}
To prove $(a)$ we need to show that  $L_{x_i}(\mathcal{R}^{(i)}_{y})=L_{x_i}(\mathcal{R}^{(i)}_{y},W_1)$, that is, 
%\begin{equation}\label{tr0}\nonumber
\[\text{rank}(\mathbf{T}(\{L_{{x_i}},x_i\},\{\mathcal{R}^{(i)}_{y},W_1\}))=  \text{rank}(\mathbf{T}(L_{{x_i}},\{\mathcal{R}^{(i)}_{y},W_1\})).\]
%\end{equation}

Since the matrix $\mathbf{T}(\mathcal{R}^{(i)}_{x},\mathcal{R}^{(i)}_{y})$ is full rank, the row  $\mathbf{T}(x_i,\mathcal{R}^{(i)}_y)$ belongs in the span of this matrix, and thus
there exist nonzero coefficients $\{\alpha_j\}$ in $\mathbf{F}_q$ such that
\begin{equation}\label{eq_a11}
\mathbf{T}(x_i,\mathcal{R}^{(i)}_y)=\sum_{x_j\in L_{x_i}(\mathcal{R}^{(i)}_{y}) }\alpha_j \mathbf{T}(x_j,\mathcal{R}^{(i)}_y).
\end{equation}
Note that for each $y_j\in W_1$, there also exist nonzero coefficients $\{\beta_j\}$ in $\mathbf{F}_q$ such that 
\begin{equation}
\text{rank}(\mathbf{T}(\{\mathcal{R}^{(i)}_{x},x_i\}, \{\mathcal{R}^{(i)}_y,y_j\})=
\text{rank}(\mathbf{T}(\mathcal{R}^{(i)}_{x}, \{\mathcal{R}^{(i)}_y,y_j\})
\end{equation}
otherwise the node $A(y_j)$ would have been visited and marked and $y_j\notin W_1$. 
Thus $\mathbf{T}(x_i,\{\mathcal{R}^{(i)}_y,W_1\})$ belongs in the span of $\mathbf{T}(\mathcal{R}^{(i)}_x,\{\mathcal{R}^{(i)}_y,W_1\})$, and
\begin{equation}\label{eq_a22}
\mathbf{T}(x_i,\{\mathcal{R}^{(i)}_y,W_1\})=\sum_{x_j\in L_{x_i}(\mathcal{R}^{(i)}_{y},W_1) }\beta_j \mathbf{T}(x_j,\{\mathcal{R}^{(i)}_y,W_1\}).
\end{equation}
Expurgating from both sides of 
(\ref{eq_a22}) the columns of $W_1$ results in an equation that still holds for the expurgated vectors
and has only columns corresponding to $\mathcal{R}^{(i)}_y$.
 From LI of all vectors $\mathbf{T}(x,\mathcal{R}^{(i)}_y)$, $x\in \mathcal{R}^{(i)}_x$,
% $\mathbf{x}_i(U_{y})$, 
   none of these expurgated vectors is identically zero. 
Moreover, from minimality of $L_{x_i}(\mathcal{R}^{(i)}_{y})$ the expansion (\ref{eq_a11}) is unique.
We thus conclude that $\alpha_j=\beta_j$ and $L_{x_i}=L_{x_i}(\mathcal{R}^{(i)}_{y})=L_{x_i}(\mathcal{R}^{(i)}_{y},W_1)$.\\
{ $(b):\quad $}
We will now argue that $L_{x_i}(\mathcal{R}^{(i)}_y)=L_{{x_i}}(\{\mathcal{R}^{(i)}_y,W_L\})$.
Consider a specific $y_k \in W_L$ that has a nonzero value in a row $x_k\in L_{x_i}$. That is, there exists an edge ($x_k,y_k$) with  $x_k\in L_{x_i}$ and  $y_k \in W_L$.

During the algorithm, we will at some point ``release'' $x_k$ from the set of used edges and replace it with $x_i$.  We will then attempt to explore $x_k$, 
assuming the set of used edges $\mathcal{R}^{(i)}$.  Note that $x_k$ might have already been explored before using a different set of 
used edges $\mathcal{R}^{(k)}$. However, our algorithm will for each $x_i$ explore all inputs in the set $L_{x_i}$ using  $\mathcal{R}^{(i)}$ again, 
even though these might have been explored before. 

If the matrix  $\mathbf{T}(\{\mathcal{R}^{(i)}_x,x_i\},\{\mathcal{R}^{(i)}_y,y_k\})$ is full rank, then the node $A(y_k)$ will be visited and $y_k \notin U_{By}$ which is a contradiction. 
Thus the matrix  $\mathbf{T}(\{\mathcal{R}^{(i)}_x,x_i\},\{ \mathcal{R}^{(i)}_y,y_k\})$ is not full rank.
 Consider then the  set $L_{x_i}(\{\mathcal{R}^{(i)}_y,y_k\})$.
Applying a similar argument as in $(a)$, we have that 
\begin{equation}\label{eq_a1}
\mathbf{T}(x_i,\mathcal{R}^{(i)}_y)=\sum_{x_j\in L_{x_i}(\mathcal{R}^{(i)}_{y}) }\alpha_j \mathbf{T}(x_i,\mathcal{R}^{(i)}_y)
\end{equation}
and
\begin{equation}\label{eq_a2}
\mathbf{T}(x_i,\{\mathcal{R}^{(i)}_y,y_k\})=\sum_{x_j\in L_{x_k}(\mathcal{R}^{(i)}_{y},y_k) }\beta_j \mathbf{T}(x_k,\{\mathcal{R}^{(i)}_y,y_k\}).
\end{equation}
Expurgating the column corresponding to $y_k$ we conclude that $L_{x_i}(\mathcal{R}^{(i)}_y\cup \{y_k\})=L_{x_i}(\mathcal{R}^{(i)}_y)$.
Repeating for all  $y_k \in W_L$ concludes $(b)$.
\\
{\em $(c):\quad$}
Clearly it also holds that 
\begin{equation}\label{tr4}\nonumber
\text{rank}(\mathbf{T}(\{L_{{x_i}},x_i\},\mathcal{R}^{(i)}_{y}))=  \text{rank}(\mathbf{T}(L_{{x_i}},\{\mathcal{R}^{(i)}_{y},W_0\})),
\end{equation}
which concludes the proof of this lemma.
%\end{itemize}
%\hfill{$\square$}
\end{proof}

%===============================================================================

\begin{lemma} \label{lemma_s11}
For each $x_i \in U_{Bx}-U_{Bx}'$, the vector
$\mathbf{T} (x_i, \{\mathcal{U}^{(i)}_y, U_{By}-U_{By}'\})$  
belongs in the span of the matrix 
$\mathbf{T} (U_{Bx}', \{\mathcal{U}^{(i)}_y, U_{By}-U_{By}'\})$,
where $\mathcal{U}^{(i)}_y$ denotes the set of unmarked  outputs in the set $U_y$. 
\end{lemma}
%====================================================================================
\begin{proof}
Note that all unmarked outputs in $U_y$  are included in $\mathcal{R}^{(i)}_{y}$,
and thus, $\mathcal{U}^{(i)}_y\subseteq \mathcal{R}^{(i)}_{y}\cap U_y$.

Order the inputs $x_i \in U_{Bx}-U_{Bx}'$ according to the order with which they are for the first time visited.
That is, $x_1$ is the first unused input that is explored inside layer $i$ and during iteration $K+1$, $x_2$ the second one, etc.
We will prove our claim through induction.\\
{\em Induction Step 1:}
When $x_1$, the first input, gets visited, clearly $\mathcal{R}^{(1)}=U$, and $\mathcal{U}_y^{(1)}=U_y$ since to perform a rewiring using a new output, we need to have already explored at least one input. 
From lemma~\ref{lemma_s01} 
we know that the  vector $\mathbf{T} (x_1, \{U_y, U_{By}-U_{By}'\})$  belongs in the span of the matrix $\mathbf{T} (U_x, \{U_y, U_{By}-U_{By}'\})$
and in particular from (\ref{eq_L_imp})
belongs in the span of the matrix  $\mathbf{T} (L_{x_1}, \{U_y, U_{By}-U_{By}'\})$. 

It is then sufficient to prove that the inputs in $L_{x_1}$ belong in  marked nodes, i.e., $L_{x_1} \subseteq U_{Bx}'$. But this holds, because of the  algorithm steps when we visit $x_1$. In particular, when $x_1$ is explored, all nodes  with $x\in L_{x_1}$ are visited, 
 marked, and explored assuming the set of used edges $U$. Thus  $L_{x_1} \subseteq U_{Bx}'$, and 
\begin{equation}
\begin{aligned}
& \text{rank}(\mathbf{T}(\{U_{Bx}',x_1\},\{U_{y},U_{By}-U'_{By}\}) = \text{rank}(\mathbf{T}(U_{Bx}',\{U_{y},U_{By}-U'_{By}\})
\end{aligned}
\end{equation}
{\em Induction Step k:} Assume that for $1\leq i \leq k$ $\mathbf{T} (x_i, \{\mathcal{U}^{(i)}_y, U_{By}-U_{By}'\})$  
belongs in the span of the matrix 
$\mathbf{T} (U_{Bx}', \{\mathcal{U}^{(i)}_y, U_{By}-U_{By}'\})$.\\
{\em Induction Step k+1:}
From lemma~\ref{lemma_s01}, we know that the vector
$\mathbf{T}( x_{k+1}, \{ \mathcal{R}_y^{(k+1)}, U_{By}-U_{By}'\})$
belongs in the span of the matrix
$\mathbf{T}(\mathcal{R}^{(k+1)}_{x}, \{ \mathcal{R}_y^{(k+1)}, U_{By}-U_{By}'\}) $,
and in particular in the span of the matrix 
$\mathbf{T}(L^{(k+1)}_{x}, \{ \mathcal{R}_y^{(k+1)}, U_{By}-U_{By}'\}) $,
where $L^{(k+1)}_{x} \subseteq \mathcal{R}^{(k+1)}_{x}$. 
Removing the columns that are not in $\mathcal{U}^{(i)}_y$, we get that 
the row $\mathbf{T}( x_{k+1}, \{ \mathcal{U}^{(i)}_y, U_{By}-U_{By}'\})$
 is in the span of the rows  $\mathbf{T}(L^{(k+1)}_{x}, \{ \mathcal{U}^{(i)}_y, U_{By}-U_{By}'\}) $.
Now, all $x \in L^{(k+1)}_{x}$ are visited and marked during the algorithm.
For each such $x$, if $x\in U_x$, then $x$ will appear in $U'_{Bx}$. 
If on the other hand $x\in \mathcal{R}^{(k+1)}_{x}$  but $x\notin U_x$,
then $x$ is one of $\{x_1,\ldots,x_k\}$
since $\mathcal{R}^{(k+1)}_{x}$ can only differ from $U_x$ on marked inputs.
From induction, for each $x_i \in \{x_1,\ldots,x_k\}$ the row vector
$\mathbf{T} (x_i, \{\mathcal{U}^{(i)}_y, U_{By}-U_{By}'\})$  
belongs in the span of the matrix 
$\mathbf{T} (U_{Bx}', \{\mathcal{U}^{(i)}_y\cap U_y, U_{By}-U_{By}'\})$.
Moreover, $\mathcal{U}^{(k+1)}_y\subseteq \mathcal{U}^{(i)}_y$, $i<k+1$,
since, if some outputs are unmarked during iteration $K+1$, 
they also are unmarked during the previous iterations. 
This concludes this proof.
\end{proof}

%===================================================================================================
%The following lemma is not needed in the proof, but is an interesting direct consequence.
\begin{lemma} \label{lemma_s1}
In the matrix   $\mathbf{T} (U_{Bx}, U_{By})$ 
each row corresponding to unused marked inputs, i.e., $x_i\in U_{Bx}-U_{Bx}'$, is in the span of the rows corresponding to inputs in $U'_{Bx}$, and thus
$ \text{rank} \mathbf{T} (U_{Bx}, U_{By})=\text{rank}  \mathbf{T} (U_{Bx}', U_{By}).$
\end{lemma}
%========================================================================================================
\begin{proof}
From Lemma~\ref{lemma_s11}, for each  $x_i\in U_{Bx}-U_{Bx}'$, we know that the row vector
$\mathbf{T} (x_i, \{\mathcal{U}^{(i)}_y, U_{By}-U_{By}'\})$  belongs in the span of the matrix 
$\mathbf{T} (U_{Bx}', \{\mathcal{U}^{(i)}_y, U_{By}-U_{By}'\})$. 
That is,
\begin{equation}\label{eq_rem}
\mathbf{T} (x_i, \{\mathcal{U}^{(i)}_y, U_{By}-U_{By}'\})=\sum_{x_j \in U_{Bx}'} \alpha_j \mathbf{T} (x_j, \{\mathcal{U}^{(i)}_y, U_{By}-U_{By}'\})
\end{equation}
for some $\alpha_j\in \mathbf{F}_q$.
Next, note that 
$U_{By}'$ is a subset of $\mathcal{U}^{(i)}_y$ for each $i$. This is because, at each rewiring, 
$\mathcal{U}^{(i)}_y$  can differ from $U_y$ only on {\em marked} outputs. 
But
$U_{By}'$ is the set of used and {\em unmarked} outputs, and thus $U_{By}'\subseteq \mathcal{U}^{(i)}_y $.
Removing some columns from both sides of (\ref{eq_rem})  we get that
\begin{equation}\nonumber
\mathbf{T} (x_i, U_{By})=
\mathbf{T} (x_i, \{U_{By}', U_{By}-U_{By}'\})=\sum_{x_j \in U_{Bx}'} \alpha_j \mathbf{T} (x_j, \{U_{By}', U_{By}-U_{By}'\})=\sum_{x_j \in U_{Bx}'} \alpha_j \mathbf{T} (x_j, U_{By})
\end{equation}
and the claim follows.
\end{proof}

%======================================================================================================

%======================================================================================================

%======================================================================================================

\begin{lemma}
\label{lemma_s3} The rank of the matrix  $\mathbf{T} (U_{Bx}, U_{By})$ can be upper bounded as
 \[ \text{rank} \mathbf{T} (U_{Bx}', U_{By})\leq |U_{Bx}'|-(|U_y|-|U_{By}'|).\]
\end{lemma}
%========================================================================================================
\begin{proof}
Consider the matrix $\mathbf{A}\triangleq \mathbf{T} (U_{Bx}', \{U_y,U_{By}-U'_{By}\})$. 
This matrix has less rows than $\mathbf{T} (U_{Bx}, U_{By})$  as it does not contain the rows in $U_{Bx}-U'_{Bx}$,
and has more columns than  $\mathbf{T} (U_{Bx}, U_{By})$ as it contains the additional columns corresponding to the 
outputs  $U_y-U_{By}'$.
The idea in this proof is to gradually change matrix $\mathbf{A}$, by sequentially adding rows and by removing columns, until we create the matrix $\mathbf{T} (U_{Bx}, U_{By})$, taking into account how each operation affects the  rank.

Order the marked outputs in $U_y$, i.e., the outputs in $U_y-U'_{By}$ that we need remove,
according to the time they got marked, i.e., $y_1$ is the output that got marked first when node $A(y_1)$ is visited, 
$y_2$ the one that got marked second, etc.  
Now, assume that    at the time when output $y_1$ is visited, $j_1$ unused inputs (not in $U_x$) 
have already been visited and marked (note that $j_1\geq 1$, if $j_1=0$ it is not possible to mark $y_1$).
In general, when output $y_k$ is visited, we will have that $j_k$ inputs in $U_{Bx}-U'_{Bx}$ are marked,
with $j_1 \leq j_2\leq j_3\ldots \leq j_L$ and $L\triangleq |U_{Bx}|-|U'_{Bx}|$.

Our starting point is that  the matrix  $\mathbf{A}$ has rank
 $|U_{Bx}'|$, i.e.,  all its rows are linearly independent.  Indeed,
since the $ K \times K$ matrix $\mathbf{T} (U_{x}, U_y)$ is full rank and $U_{Bx}'\subseteq U_x$,  
the rows  $\mathbf{T} (U_{Bx}', U_y)$ are LI, and as a result so are the rows
$\mathbf{T} (U_{Bx}', \{U_y,U_{By}-U'_{By}\})$.

We are going to perform $L=|U_y|-|U_{By}'|$ steps, 
creating a sequence of matrices $\{ \mathbf{A}_0=\mathbf{A}$, $\mathbf{A}_1,\ldots,\mathbf{A}_L \}$ where at step $k$, $k=1,\ldots L$,  we first add to matrix $\mathbf{A}_{k-1}$ the rows $\{ x_{j_{k-1}+1} \ldots x_{j_k}\}$
 and then we remove the output $y_k$ in $(U_y-U_{By}')$ to create the matrix $\mathbf{A}_{k}$.\\
{\em Step 1: Removing output $y_1$.}\\  
Let $\mathcal{R}^{(j_1)}$ be the set of perceived used edges when $y_1$ is marked.
Since this is the first time an output in $U_y$ is marked and the $\phi$ function is executed, 
$\mathcal{R}^{(j)}_y= U_y$, for all $j\leq j_1$.

We know from lemma~\ref{lemma_s11} that   the rows $\mathbf{T} (x_i, \{U_y, U_{By}-U'_{By}\})$, 
$1\leq i \leq j_1$ belong in the span of the matrix  $\mathbf{T} (U'_{Bx}, \{U_y, U_{By}-U'_{By}\})$.
Thus adding these rows to matrix $\mathbf{A}$ does not increase its rank.

From lemma~\ref{lemma_extra},  there exist a set of rows $S(y_1)$ with
$S(y_1)\subseteq \{ \mathcal{R}^{(j_1)}_x,x_1,\ldots,x_{j_1}\}$
such that, removing the column $y_1$
drops the rank of the matrix $\mathbf{T} (S(y_1), \{\mathcal{R}^{(j_1)}_y, U_{By}-U'_{By}\})$
from $|S(y_1)|$ to $|S(y_1)|-1$.
In other words, the column  $\mathbf{T} (S(y_1),y_1)$ is LI from all the  columns of the matrix 
$\mathbf{T} (S(y_1), \{\mathcal{R}^{(j_1)}_y-y_1, U_{By}-U'_{By}\})$.

Notice that when the node $A(y_1)$ gets visited during iteration $K+1$, we will execute the  $\phi$-function for output $y_1$.
As a result, all the nodes with inputs in   $S(y_1)$ will be visited and marked by the algorithm during iteration $K+1$.  
Thus we know that $S(y_1)\subset\{U'_{Bx},x_1,\ldots,x_{j_1}\}$, that is, they form part of the set of  marked inputs by the algorithm.

Since the ``partial'' column  $\mathbf{T} (S(y_1),y_1)$ is LI from the columns in the matrix  $\mathbf{T} (S(y_1), \{U_y-y_1, U_{By}-U'_{By}\})$, it
 follows immediately that the column $\mathbf{T} (\{U'_{Bx},x_1,\ldots x_{j_1}\},y_1)$ is LI from the columns in the matrix
$\mathbf{T} (\{U'_{Bx},x_1,\ldots x_{j_1}\}, \{\mathcal{R}^{(j_1)}_y-y_1, U_{By}-U'_{By}\})$.
Thus if we drop the column $y_1$ from the matrix $\mathbf{T} (\{U'_{Bx},x_1,\ldots x_{j_1}\}, \{U_y, U_{By}-U'_{By}\})$ the resulting matrix
$\mathbf{A}_1\triangleq \mathbf{T} (\{U'_{Bx},x_1,\ldots x_{j_1}\}, \{\mathcal{R}^{(j_1)}_y-y_1, U_{By}-U'_{By}\})$ has rank $|U_{Bx}'|-1$.\\
{\em Step k: Removing output $y_k$.}\\
We start from the matrix $\mathbf{A}_{k-1}\triangleq \mathbf{T} (\{U'_{Bx},x_1,\ldots x_{j_{k-1}}\}, \{\mathcal{R}^{(j_1)}_y-y_1-\ldots-y_{k-1}, U_{By}-U'_{By}\})$ that has rank $|U_{Bx}'|-(k-1)$.
From lemma~\ref{lemma_s11} the rows $\mathbf{T} (x_j, \{U_y-y_1-\ldots-y_{k-1}, U_{By}-U'_{By}\})$, 
$j_{k-1} \leq j \leq j_k$ belong in the span of the matrix  $\mathbf{T} (U'_{Bx}, \{U_y-y_1-\ldots-y_{k-1}, U_{By}-U'_{By}\})$.
Thus adding these rows to matrix $\mathbf{A}_{k-1}$ does not increase its rank.

On the other hand, from lemma~\ref{lemma_extra}
 there exists a set of LI rows  $S(y_k)\subseteq  \{ \mathcal{R}^{(j_k)}_x,x_1,\ldots x_{j_{k-1}}\}$
such that removing the column $y_k$ from the matrix  $\mathbf{T} (S(y_k), \{\mathcal{R}^{(j_k)}_y, U_{By}-U'_{By}\})$
drops the rank of this matrix
from $|S(y_k)|$ to $|S(y_k)|-1$.
In other words, the column  $\mathbf{T} (S(y_k),y_k)$ is LI from all the  columns of the matrix 
$\mathbf{T} (S(y_k), \{\mathcal{R}^{(j_k)}_y-y_k, U_{By}-U'_{By}\})$.
But  $\mathcal{R}^{(j_k)}_y$ contains all the outputs in 
$U_y-y_1-\ldots-y_{k-1}$, and thus the column $\mathbf{T} (S(y_k), y_k)$ does not belong in the span of the columns
$\mathbf{T} (S(y_k), \{U_y-y_1-\ldots-y_{k-1}, U_{By}-U'_{By}\})$.

Similar to before because the $\phi$-function will be executed at $y_k$, all the inputs in $S(y_k)$ are marked and
 $S(y_k)\subset\{U'_{Bx},x_1,\ldots,x_{j_k}\}$.
It again
 follows immediately that the column $\mathbf{T} (\{U'_{Bx},x_1,\ldots x_{j_k}\},y_k)$ is LI from the columns in the matrix
$\mathbf{T} (\{U'_{Bx},x_1,\ldots x_{j_k}\}, \{\mathcal{R}^{(j_1)}_y-y_1-\ldots-y_k, U_{By}-U'_{By}\})$.
Thus if we drop the column $y_k$ from the matrix $\mathbf{T} (\{U'_{Bx},x_1,\ldots x_{j_k}\}, \{U_y-y_1-\ldots-y_{k-1}, U_{By}-U'_{By}\})$ the resulting matrix
$\mathbf{A}_k\triangleq \mathbf{T} (\{U'_{Bx},x_1,\ldots x_{j_k}\}, \{\mathcal{R}^{(j_1)}_y-y_1-\ldots-y_k, U_{By}-U'_{By}\})$ has rank $|U_{Bx}'|-k$.\\
{\em  Final step.}\\
At the end of this procedure, the matrix $\mathbf{A}_L\triangleq \mathbf{T} (\{U'_{Bx},x_1,\ldots x_{j_L}\}, \{\mathcal{R}^{(j_1)}_y-y_1-\ldots-y_L, U_{By}-U'_{By}\})$ has rank $|U_{Bx}'|-L$ and the required column set $U_{By}$. Now to create the matrix  $\mathbf{T} (U_{Bx},U_{By})$ we may need to add to $\mathbf{A}_L$ some additional rows. From Lemma~\ref{lemma_s1} adding these rows cannot increase the rank of the matrix as they belong in the span of $\mathbf{T} (U'_{Bx},U_{By})$. This completes our proof.
\end{proof}
   
%================================================================================================
\begin{lemma}\label{lemma_extra}
Let $j$ denote the number of already marked inputs when output $y$ gets marked, and let $\mathcal{R}^{(j)}_y$ denote the preceived set of used outputs from previous iterations at that time. 
Then there exists a  set of rows $S(y)$ in the set $\{\mathcal{R}^{(j)}_x,x_1,\ldots,x_j\}$ such that
\begin{equation} \label{toprove}
\begin{aligned}
 &\text{rank}\mathbf{T} (S(y), \{\mathcal{R}^{(j)}_y, U_{By}-U'_{By}\})=|S(y)|, \mbox{ while } \\
&\text{rank}\mathbf{T} (S(y), \{\mathcal{R}^{(j)}_y-y, U_{By}-U'_{By}\})=|S(y)|-1.
\end{aligned}
\end{equation}
That is, removing the column  $y$ drops the rank of the matrix by one, and makes the rows $S(y)$ LD.
\end{lemma}
%==================================================================================================
\begin{proof}
Consider iteration $K+1$ and layer $i$.
Assume that the node where an output $y$ in $U_y$ belongs gets visited for the first time. This can happen in two ways:
\begin{itemize}
\item {\em Case 1:} The node $A(y)$ gets visited while we perform an $Lx$-function at layer $i+1$ (see examples~\ref{example_3} and \ref{example_4}). Note that sice we have arrived at layer $i+1$, we have identified at layer $i$ an edge $(x',y')$ that is LI from the $K$ edges identified from previous iterations.
\item {\em Case 2:} The node $A(y)$ gets visited when we find an edge $(x',y')$ in layer $i$ with 
$\text{rank}\mathbf{T} (\{\mathcal{R}^{(j)}_x,x'\}, \{\mathcal{R}^{(j)}_y),y'\}=K+1$ (see examples~\ref{example_2} and \ref{example_3}).
\end{itemize}
The arguments in these two cases are very similar, and we treat them together.
In both cases, at layer $i$, we start with the $(K+1)\times (K+1)$ full rank matrix
 $\mathbf{T} (\{\mathcal{R}^{(j)}_x,x'\}, \{\mathcal{R}^{(j)}_y,y'\}).$ 
%.\\
%{\em Case 1.}\\
%Consider the full rank  $K\times K$ matrix $\mathbf{T} (\mathcal{R}^{(j)}_x, \mathcal{R}^{(j)}_y)$. 
When we remove the column $y$ 
clearly the resulting $(K+1)\times K$ matrix has some linearly dependent rows.
As a result, a subset of the rows becomes linearly dependent. Define 
$S (y)$  to be the set of inputs in $\{\mathcal{R}^{(j)}_x,x'\}$ corresponding to the {\em minimally linearly dependent} rows in the matrix $\mathbf{T} (\{\mathcal{R}^{(j)}_x,x'\}, \{\mathcal{R}^{(j)}_y-y,y'\})$,
where by { minimally linear dependent} we mean that the vectors in the set  are linear dependent 
but any proper subset of them is a linearly independent set of vectors. 
Note that the inputs in $S(y)$ are exactly the inputs that are going to be visited when the algorithm performs the $\phi$-function for output $y$,
as, removing any of the rows in $S(y)$ from the matrix $\mathbf{T} (\{\mathcal{R}^{(j)}_x,x'\}, \{\mathcal{R}^{(j)}_y-y,y'\})$
results in a full rank $K\times K$ submatrix.

Now, since  $\mathbf{T}( \{\mathcal{R}^{(j)}_x,x'\}, \{\mathcal{R}^{(j)}_y, y'\} )$ is a full rank matrix 
then  there is no set of rows of this matrix which are linearly dependent. In particular,  the  rows in $S(y)$ are linearly independent.
The matrix $\mathbf{T} (S(y), \{ \mathcal{R}^{(j)}_y, y',U_{By} - U_{By'}\})$ contains the full rank submatrix $\mathbf{T} (S(y), \{\mathcal{R}^{(j)}_y,y'\})$
and thus  has also rank $|S(y)|$. That is
\begin{equation} \label{peq1}
\begin{aligned}
& \text{rank}\mathbf{T}( \{\mathcal{R}^{(j)}_x,x'\}, \{\mathcal{R}^{(j)}_y,y'\})=\text{rank}\mathbf{T}( \{\mathcal{R}^{(j)}_x,x'\}, \{\mathcal{R}^{(j)}_y,y',U_{By} - U_{By'}\}) =K+1, \mbox{ and }\\
&\text{rank}\mathbf{T} (S(y), \{ \mathcal{R}^{(j)}_y, y'\})=\text{rank}\mathbf{T} (S(y), \{ \mathcal{R}^{(j)}_y, y',U_{By} - U_{By'}\})=|S(y)|.
\end{aligned}
\end{equation}
Moreover, from construction,
\begin{equation} \label{peqs0}
\begin{aligned}
& \text{rank}\mathbf{T}( \{\mathcal{R}^{(j)}_x,x'\}, \{\mathcal{R}^{(j)}_y-y,y'\}) =K, \mbox{ and }\\
& \text{rank}\mathbf{T}(S(y), \{\mathcal{R}^{(j)}_y-y,y'\} )=|S(y)|-1.
\end{aligned}
\end{equation}
We will next argue that
\begin{equation} \label{eqs00}
\begin{aligned}
& \text{rank}\mathbf{T}( \{\mathcal{R}^{(j)}_x,x'\}, \{\mathcal{R}^{(j)}_y-y,y',U_{By} - U_{By'}\} )=K, \mbox{ and }\\
& \text{rank}\mathbf{T}(S(y), \{\mathcal{R}^{(j)}_y-y,y',U_{By} - U_{By'}\} )=|S(y)|-1.
\end{aligned}
\end{equation}
that is, adding the columns  in $ U_{By} - U_{By'}$ does not increase the rank.

Let $y_0$ be a column in  $U_{By} - U_{By'}$, and consider the matrix  $\mathbf{T}( \{\mathcal{R}^{(j)}_x,x'\}, \{\mathcal{R}^{(j)}_y-y,y',y_0\} )$.
 If this square matrix has rank $K+1$,
then the rows  $\mathbf{T}(S(y), \{ \mathcal{R}^{(j)}_y-y, y',y_0\} )$ must be LI.
Since the rows $\mathbf{T}(S(y),  \{\mathcal{R}^{(j)}_y-y,y' \})$ are LD, there exists row $x_0\in S(y)$
with a nonzero value in the column $y_0$. But when we run the $\phi$-function, 
as we already mentioned, all inputs in  $S(y)$ including $x_0$ are visited and explored. Thus if the 
$\mathbf{T}( \{\mathcal{R}^{(j)}_x,x'\}, \{ \mathcal{R}^{(j)}_y-y,y', y_0\} )$ were full rank, the output $y_0$ would get marked and not appear in    $U_{By} - U_{By'}$. We conclude that for every $y_0$  in  $U_{By} - U_{By'}$, the column  $\mathbf{T}(\{\mathcal{R}^{(j)}_x,x'\}, y_0\} )$
belongs in the span of the columns  $\mathbf{T}( \{\mathcal{R}^{(j)}_x,x'\}, \{\mathcal{R}^{(j)}_y-y,y'\} )$, and thus,
the matrix   $\mathbf{T}( \{\mathcal{R}^{(j)}_x,x'\}, \{ \mathcal{R}^{(j)}_y-y,y', U_{By} - U_{By'}\} )$
has rank $K$.

Next note that, the rows in matrix $\mathbf{T} (S(y), \{\mathcal{R}^{(j)}_y-y,y'\})$
do not belong in the span of the LI rows  $\mathbf{T}( \mathcal{R}^{(j)}_x-S(y), \{\mathcal{R}^{(j)}_y-y,y'\} )$. 
Let  $y_0$ be a column in  $U_{By} - U_{By'}$.
Clearly, the  rows $\mathbf{T} (S(y), \{\mathcal{R}^{(j)}_y-y,y',y_0\})$ do not belong in the span of the LI rows 
 $\mathbf{T}( \mathcal{R}^{(j)}_x-S(y), \{\mathcal{R}^{(j)}_y-y,y',y_0\} )$.
Thus, if the rows
 $\mathbf{T} (S(y), \{\mathcal{R}^{(j)}_y-y,y', y_0\})$ were LI, the matrix $\mathbf{T}( \mathcal{R}^{(j)}_x, \{\mathcal{R}^{(j)}_y-y,y',y_0\} )$ would have rank $K+1$, which is not possible from our previous argument. We conclude that 
$\text{rank}\mathbf{T} (S(y), \{ \mathcal{R}^{(j)}_y-y,y', U_{By} - U_{By'}\})=|S(y)|-1$. We have thus proved that 
\begin{equation} \label{eqs0}
\begin{aligned}
& \text{rank}\mathbf{T}( S(y), \{\mathcal{R}^{(j)}_y,y',U_{By} - U_{By'}\} )=|S(y)|, \mbox{ while }\\
& \text{rank}\mathbf{T}(S(y), \{\mathcal{R}^{(j)}_y-y,y',U_{By} - U_{By'}\} )=|S(y)|-1.
\end{aligned}
\end{equation}
Removing the column $y'$ from the last equation, we also get  that
\begin{equation}\label{eqs1}
\text{rank}\mathbf{T}(S(y), \{\mathcal{R}^{(j)}_y-y,U_{By} - U_{By'} \})\leq |S(y)|-1.
\end{equation}
We now distiguish  two cases:
\begin{enumerate}
\item $S(y) \subseteq \mathcal{R}^{(j)}_x$, i.e., the set of rows $S(y)$ does not contain $x'$. Then 
\begin{equation}\label{eqs2}
\text{rank}\mathbf{T}(S(y), \{\mathcal{R}^{(j)}_y,U_{By} - U_{By'} \})= |S(y)|-1,
\end{equation}
since  these rows are LI, and (\ref{eqs0})-(\ref{eqs2}) imply (\ref{toprove}).
\item  $x' \in S'(y)$, we have two subscases:
\begin{enumerate}
\item  if $\text{rank}\mathbf{T}(S(y), \{\mathcal{R}^{(j)}_y,U_{By} - U_{By'}\} )=|S(y)|,$ this  together with (\ref{eqs1}) implies (\ref{toprove}).
\item  if $\text{rank}\mathbf{T}(S(y), \{\mathcal{R}^{(j)}_y,U_{By} - U_{By'}\} )=|S(y)|-1,$ then given (\ref{eqs0}) the column $y'$ does not belong in the span of the columns  $\{\mathcal{R}^{(j)}_y,U_{By} - U_{By'}\}$.  Similarly, again from (\ref{eqs0}),
the column $y$ does not belong in the span of the columns  $\{\mathcal{R}^{(j)}_y-y,y',U_{By} - U_{By'}\}$. We conclude that
$\text{rank}\mathbf{T}(S(y), \{\mathcal{R}^{(j)}_y-y,U_{By} - U_{By'}\} )=|S(y)|-2$, and thus $\text{rank}\mathbf{T}(S(y)-x', \{\mathcal{R}^{(j)}_y-y,U_{By} - U_{By'}\} )\leq|S(y)|-2$. But $\text{rank}\mathbf{T}(S(y)-x', \{\mathcal{R}^{(j)}_y,U_{By} - U_{By'}\} )=|S(y)|-1$. For the set $S'(y)=S(y)-x'$, the claim in  (\ref{toprove}) follows.
\end{enumerate}
\end{enumerate}
\end{proof}
%==============================================================================================

\subsection{Algorithm Complexity}

\begin{proposition}
  \label{prop9}The complexity of the algorithm in Table~\ref{tabl1} is $\mathcal{O}(\mathcal{C}^{5}(O_{total}+\Delta_II_{total}))$, where  $\mathcal{C}$ is the capacity of the network,  $I_{total}$ equal the total number of inputs and $O_{total}$ is the total number of outputs in the network.
\end{proposition}

 \begin{proof}
At iteration $K$, the complexity of the function ``FindL($\mathbf{T}$)'' is $\mathcal{O}(K^3)$, ``Match($\mathbf{T}$)'' is $\mathcal{O}(K^3)$, to find which inputs to visit with the $\phi$-function is $O(K^3)$,  and the rank calculations are $\mathcal{O}(K^3)$.
When we visit each input we will perform at most $\Delta_I$ rank calculations, where $\Delta_I$ is the maximum outdegree of an input. This results in complexity $\mathcal{O}(\Delta_I K^3)$. 
 Moreover, we will perform the ``FindL'' function at most once for every input. 
Performing the ``FindL'' function at the $K$ iteration might result in at most $K$ inputs to be revisited. For each of the revisited inputs, the associated complexity will be $\mathcal{O}(\Delta_I K^3)$. 
Thus, the total complexity when visiting each input is $\mathcal{O}(\Delta_I K^4)$.
These operations will be repeated 
at most $I_{total}$ times.
An upper bound  for $I_{total}$ is  $|E|$, where $E$ is the set of all edges in the network, but this bound might be very loose, if the inputs have small outdegree.
To conclude, examining the inputs results in complexity of $\mathcal{O}(\Delta_IK^4I_{total})$.
When each output is marked, we will perform exactly once the $\phi$-function. Thus, this function will be performed at most once for every output (if the output gets marked during the iteration), and contributes complexity  
$\mathcal{O}(K^4O_{total})$, where  (again, a loose upper bound for $O_{total}$ is $|E|$).  
After $\mathcal{C}$ iterations the total complexity is $\mathcal{O}(\mathcal{C}^{5}(O_{total}+\Delta_II_{total}))$.
\end{proof}

\section{Conclusions}\label{sec3}

In this paper we develop a polynomial time algorithm for unicast
connections that allow to achieve the min-cut capacity in networks of
linear deterministic channels over a finite field $\mathbf{F}_q$. 
Such networks have
recently found applicability as approximate models for wireless
Gaussian networks, by modeling broadcasting and interference through
linear operations over a finite field. Our scheme allows to identify
the min-cut value in polynomial time, and to achieve this value using
very simple one symbol mapping operations at the intermediate network
nodes.


\begin{thebibliography}{1}
  \bibitem{Menger} K. Menger, ``Zur allgemeinen Kurventheorie,''
  \tmtextit{Fund. Math.} vol.~10, pp.~95-115, 1927. 

  \bibitem{Ford} L. R.  Ford, Jr. and D. R. Fulkerson, ``Maximal flow through a network,''
  \tmtextit{Canadian Journal of Mathematics,} vol.~8, pp.~399--404, 1956.

  \bibitem{Shannon} P. Elias, A. Feinstein, and C. E. Shannon, ``Note on
  maximum flow through a network,'' \tmtextit{IRE Trans. Information Theory,}
  vol.~2, pp.~117--119, 1956.
  
  \bibitem{suhas1} S. Avestimehr, S N. Diggavi and D N C. Tse, ``Wireless
  network information flow'', \tmtextit{Proceedings of Allerton Conference on
  Communication, Control, and Computing}, Illinois, September 2007. See
  http://licos.epfl.ch/index.php?p=research\_projWNC
  
  \bibitem{suhas2} S. Avestimehr, S N. Diggavi and D N C. Tse, ``A
  deterministic approach to wireless relay networks'', \tmtextit{Proceedings
  of Allerton Conference on Communication, Control, and Computing}, Illinois,
  September 2007. See http://licos.epfl.ch/index.php?p=research\_projWNC
  
  \bibitem{suhas3} S. Avestimehr, S N. Diggavi and D N C. Tse, ``Approximate
  capacity of gaussian relay networks'', \tmtextit{IEEE Symposium on
  Information Theory (ISIT)}, Toronto, July 2008.
  
  \bibitem{Ahl} R.~Ahlswede, N.~Cai, S-Y.~R.~Li, and R.~W.~Yeung, ``Network
  information flow,'' \tmtextit{IEEE Trans. Inform. Theory,} vol.~46,
  pp.~1204--1216, July~2000.
  
  \bibitem{Jaggi} S. Jaggi, P. Sanders, P. Chou, M. Effros, S. Egner, K.
  Jain and L. Tolhuizen, ``Polynomial time algorithms for multicast network
  code construction,'' \tmtextit{IEEE Trans. Inform. Theory,} vol.~51, no.~6,
  pp.~1973--1982, 2005. 

  \bibitem{Harvey} N. Harvey, ``Deterministic network
  coding by matrix completion,'' MS Thesis, MIT 2005.
  
  \bibitem{Cover} T. Cover and J. Thomas, ``Elements of Information
  Theory'', Wiley 2006.

   \bibitem{soda} A. Amaudruz, C. Fragouli, ``Combinatorial algorithms for wireless information flow'', SODA 2009.

\bibitem{matroid_allerton} M. Goemans, S. Iwata and R. Zenklusen, ``An algorithmic framework for wireless information flow'', Allerton 2009.
\end{thebibliography}
\end{document}